\documentclass[11pt]{article}
\usepackage{graphicx}
\usepackage{amsfonts}
\usepackage{amsmath}
\usepackage{amssymb}
\usepackage[margin=2.54cm]{geometry}

\begin{document}

\title{Seismic Solvability Problems}

\author{August Lau and Chuan Yin \\
        Email contact: \texttt{chuan.yin@apachecorp.com}}

\date{December 6, 2012}

\maketitle

\begin{abstract}

Classical approach of solvability problem has shed much light on what we can solve and what we cannot solve mathematically.   Starting with quadratic equation, we know that we can solve it by the quadratic formula which uses square root.   Polynomial is a generalization of quadratic equation.   If we define solvability by using only square roots, cube roots etc, then polynomials are not solvable by radicals (square root, cube root etc).   We can classify polynomials into simple (solvable by radicals) and complex (not solvable by radicals).  We will use the same metaphor to separate what is solvable (simple part) and what is not solvable (complex part).

This paper is a result of our presentation at a University of Houston seminar. In this paper, we will study seismic complexity through the eyes of solvability.   We will investigate model complexity, data complexity and operator complexity.   Model complexity is demonstrated by multiple scattering in a complex model like Cantor layers. Data complexity is studied through Betti numbers (topology/cohomology).  Data can be decomposed as simple part and complex part.   The simple part is solvable as an inverse problem.   The complex part could be studied qualitatively by topological method like Betti numbers.  Operator complexity is viewed through semigroup theory, specifically through idempotents (opposite of group theory).  Operators that form a group are invertible (solvable) while semigroup of operators is not invertible (not solvable) in general.
\end{abstract}

\newpage

\begin{figure}
\centering
  \includegraphics[width=5.0in]{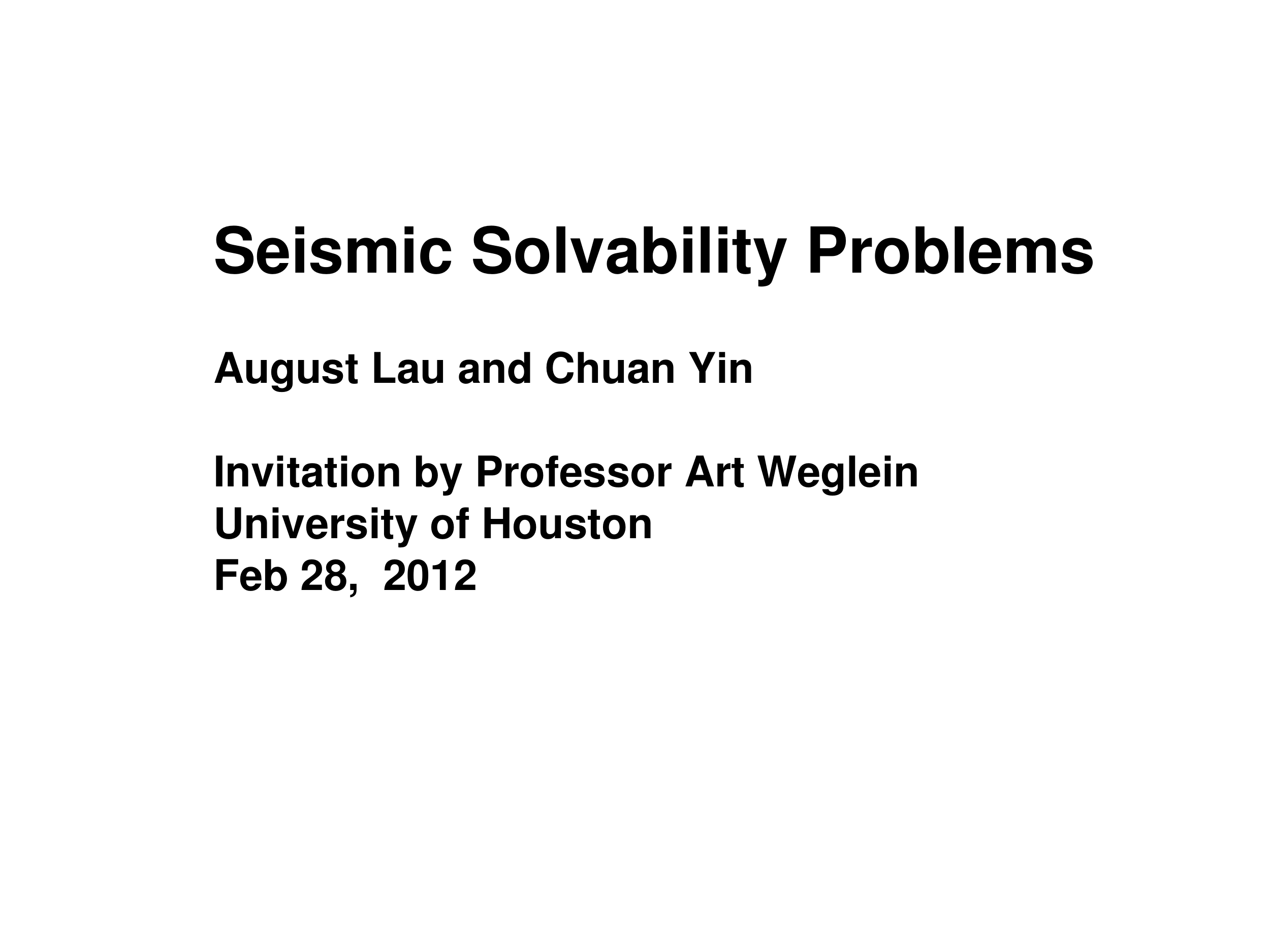}
\caption{Slide 1}
\end{figure}

\begin{figure}
\centering
  \includegraphics[width=5.0in]{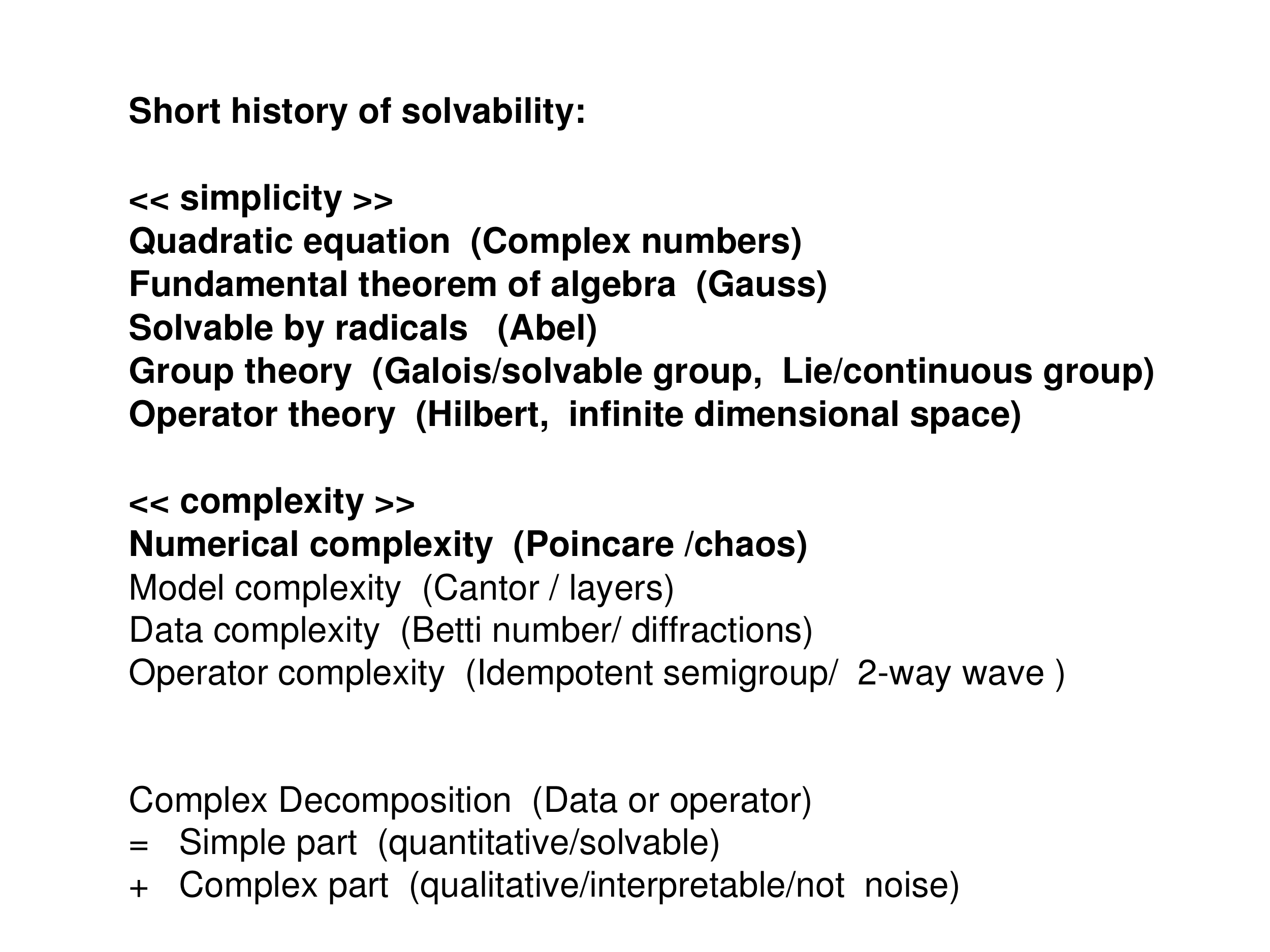}
\caption{Slide 2}
\end{figure}

\begin{figure}
\centering
  \includegraphics[width=5.0in]{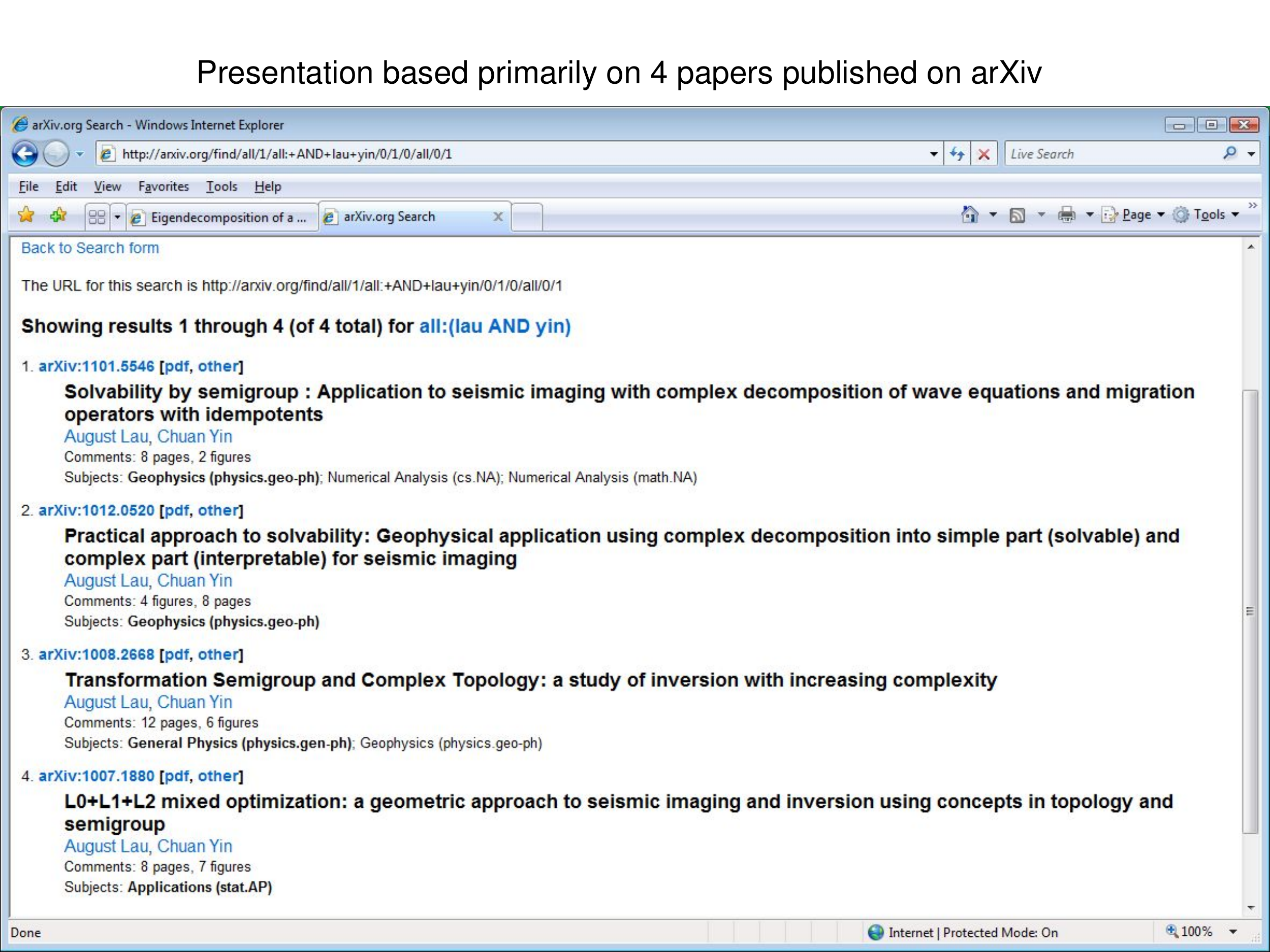}
\caption{Slide 3}
\end{figure}

\begin{figure}
\centering
  \includegraphics[width=5.0in]{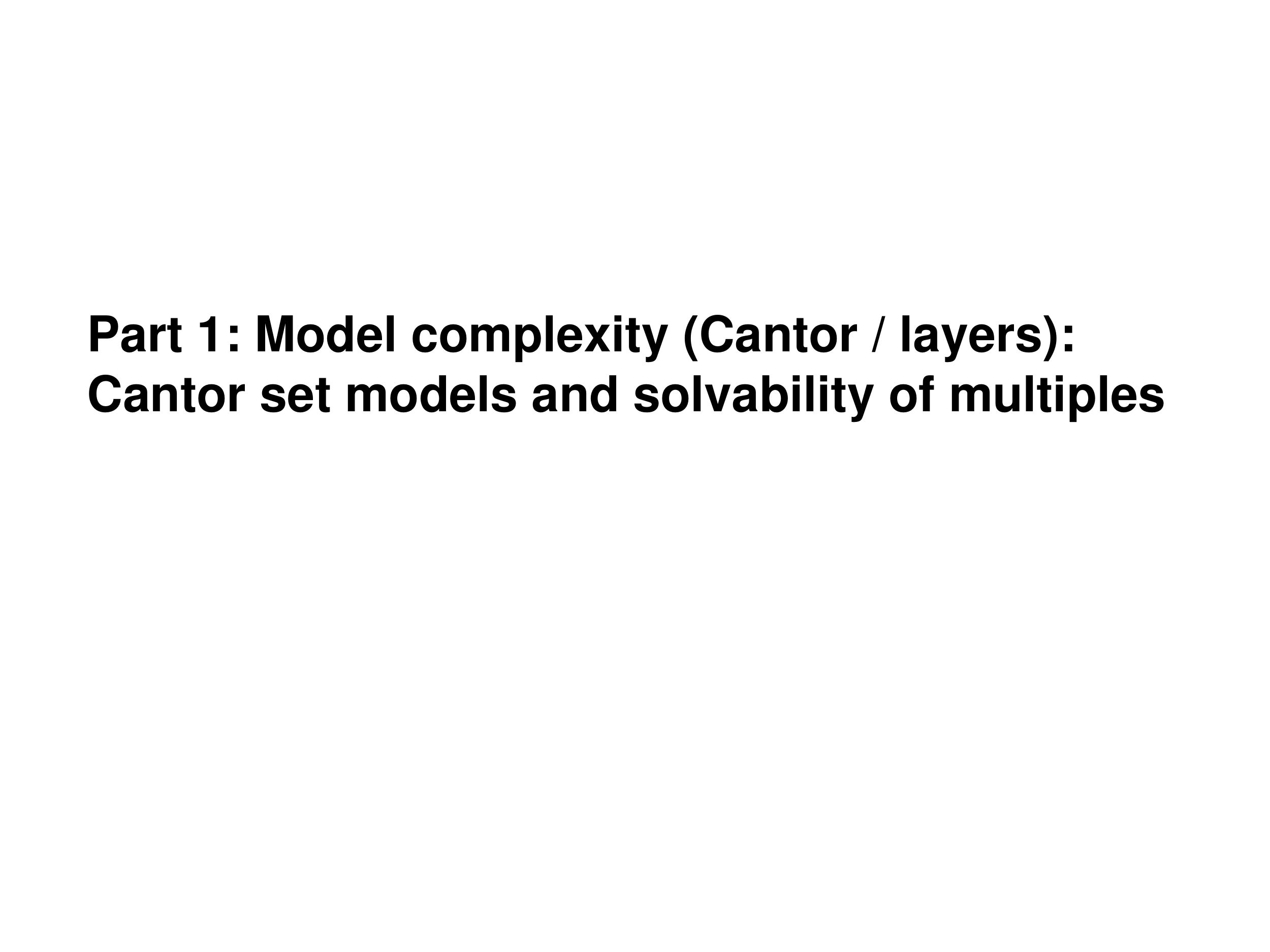}
\caption{Slide 4}
\end{figure}

\begin{figure}
\centering
  \includegraphics[width=5.0in]{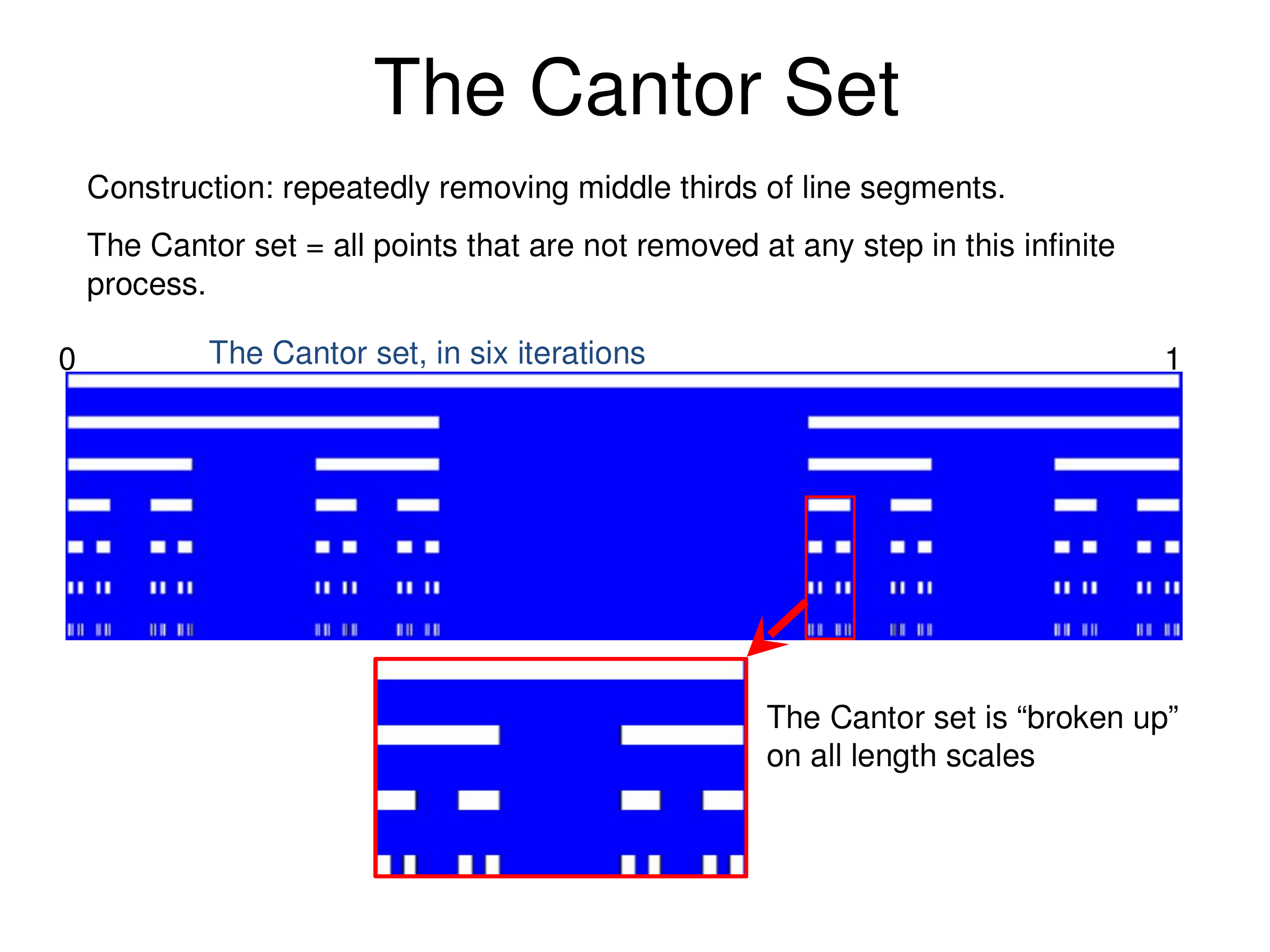}
\caption{Slide 5}
\end{figure}

\begin{figure}
\centering
  \includegraphics[width=5.0in]{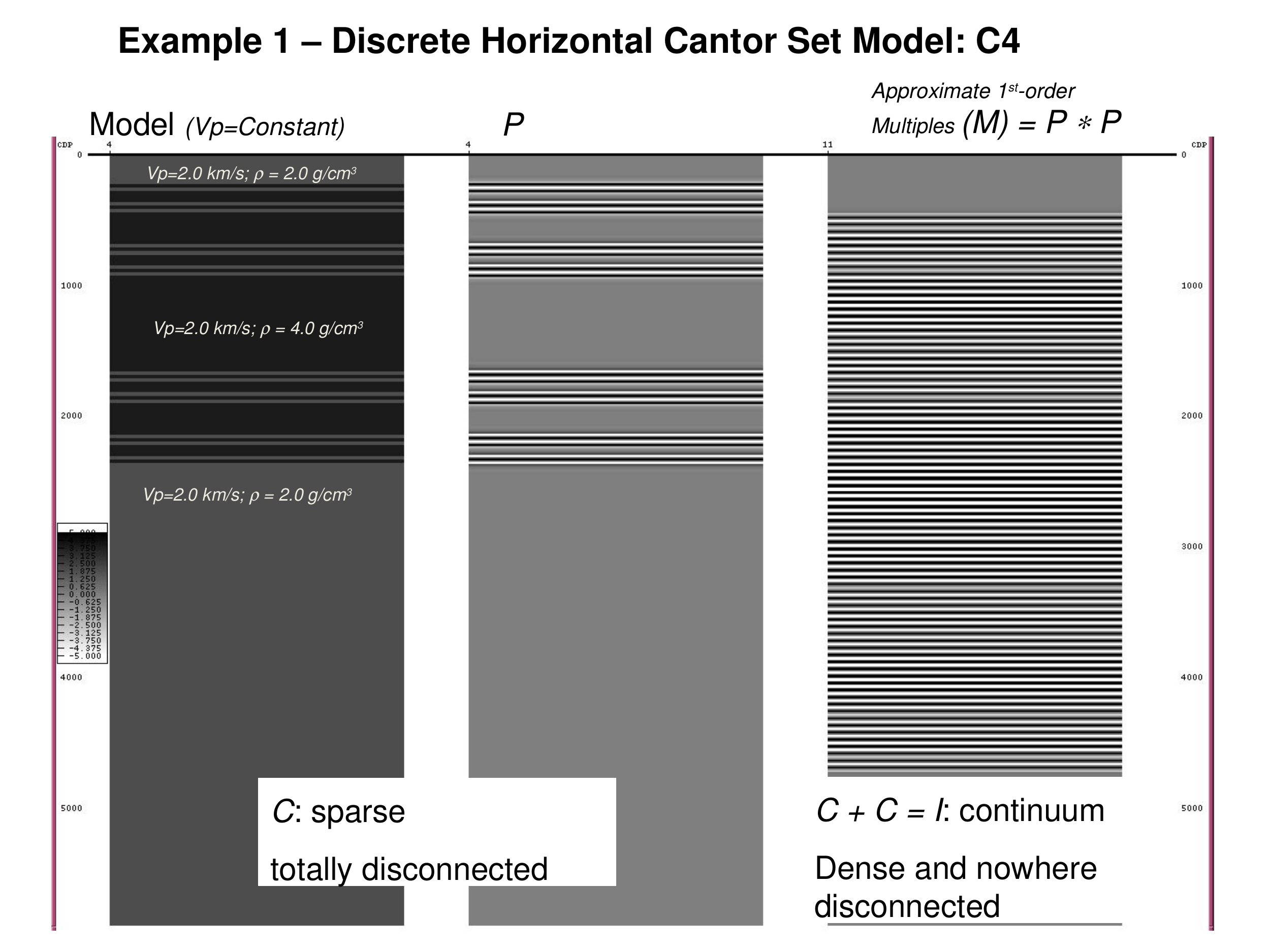}
\caption{Slide 6}
\end{figure}

\begin{figure}
\centering
  \includegraphics[width=5.0in]{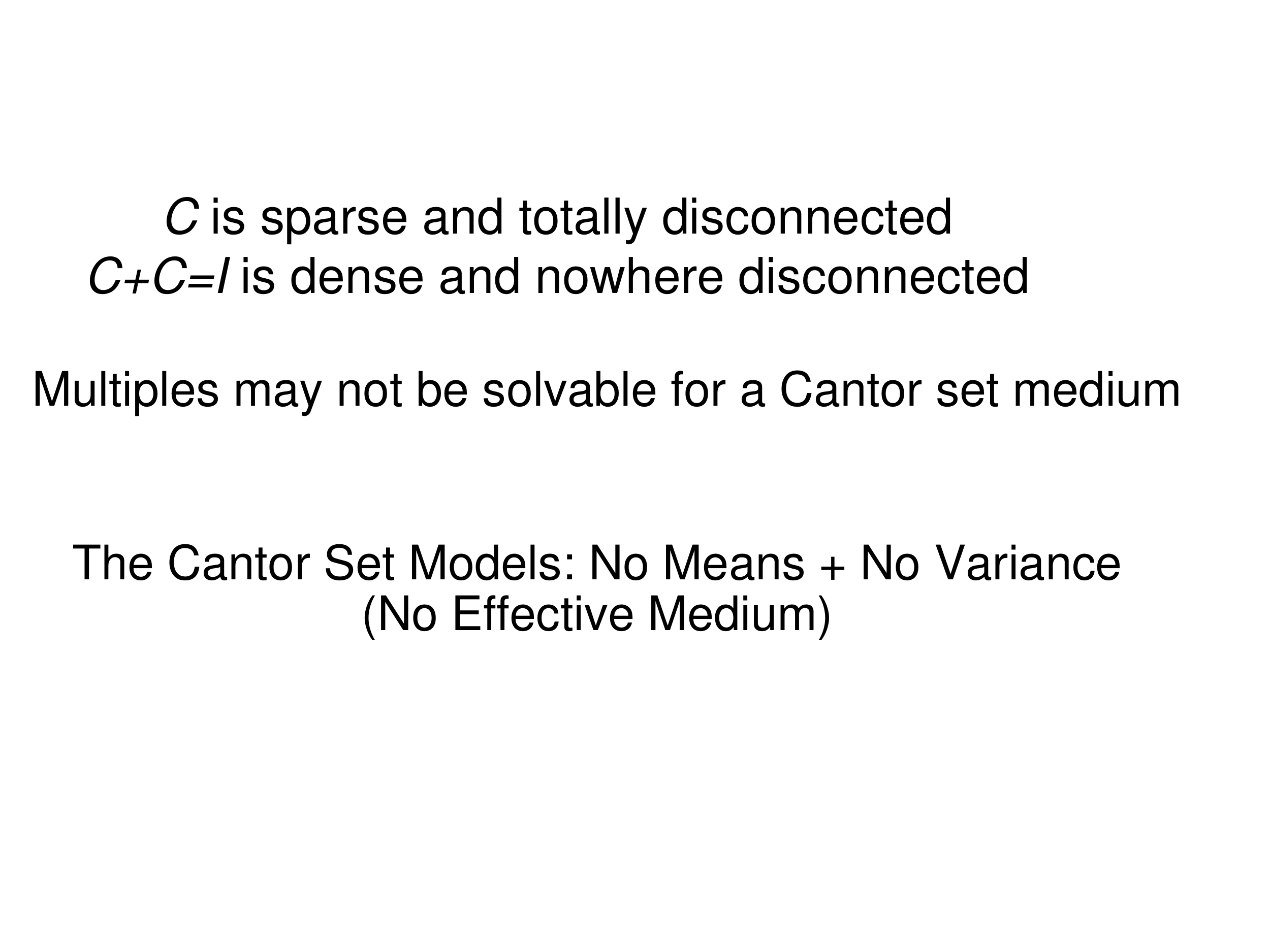}
\caption{Slide 7}
\end{figure}

\begin{figure}
\centering
  \includegraphics[width=5.0in]{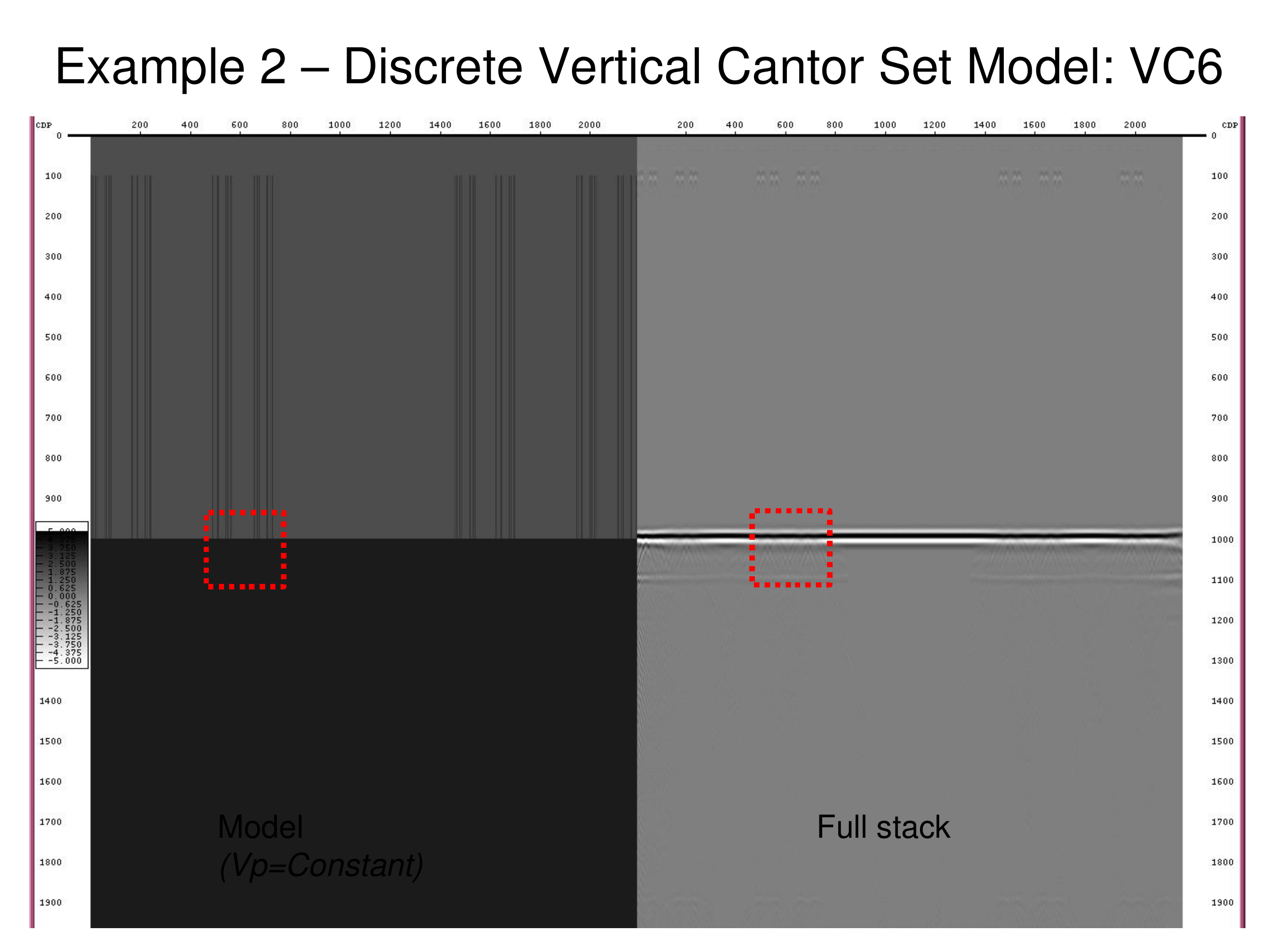}
\caption{Slide 8}
\end{figure}

\begin{figure}
\centering
  \includegraphics[width=5.0in]{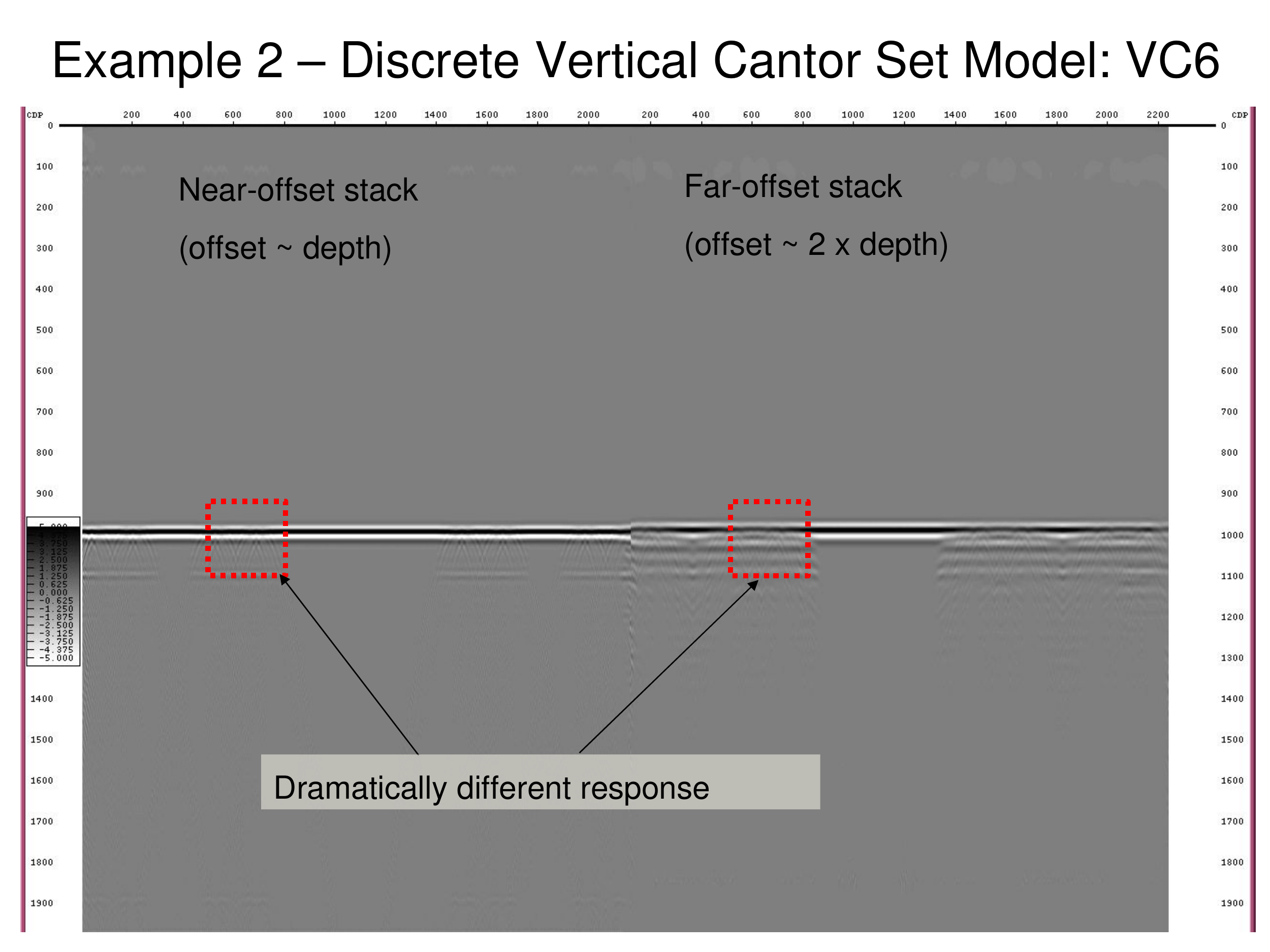}
\caption{Slide 9}
\end{figure}

\begin{figure}
\centering
  \includegraphics[width=5.0in]{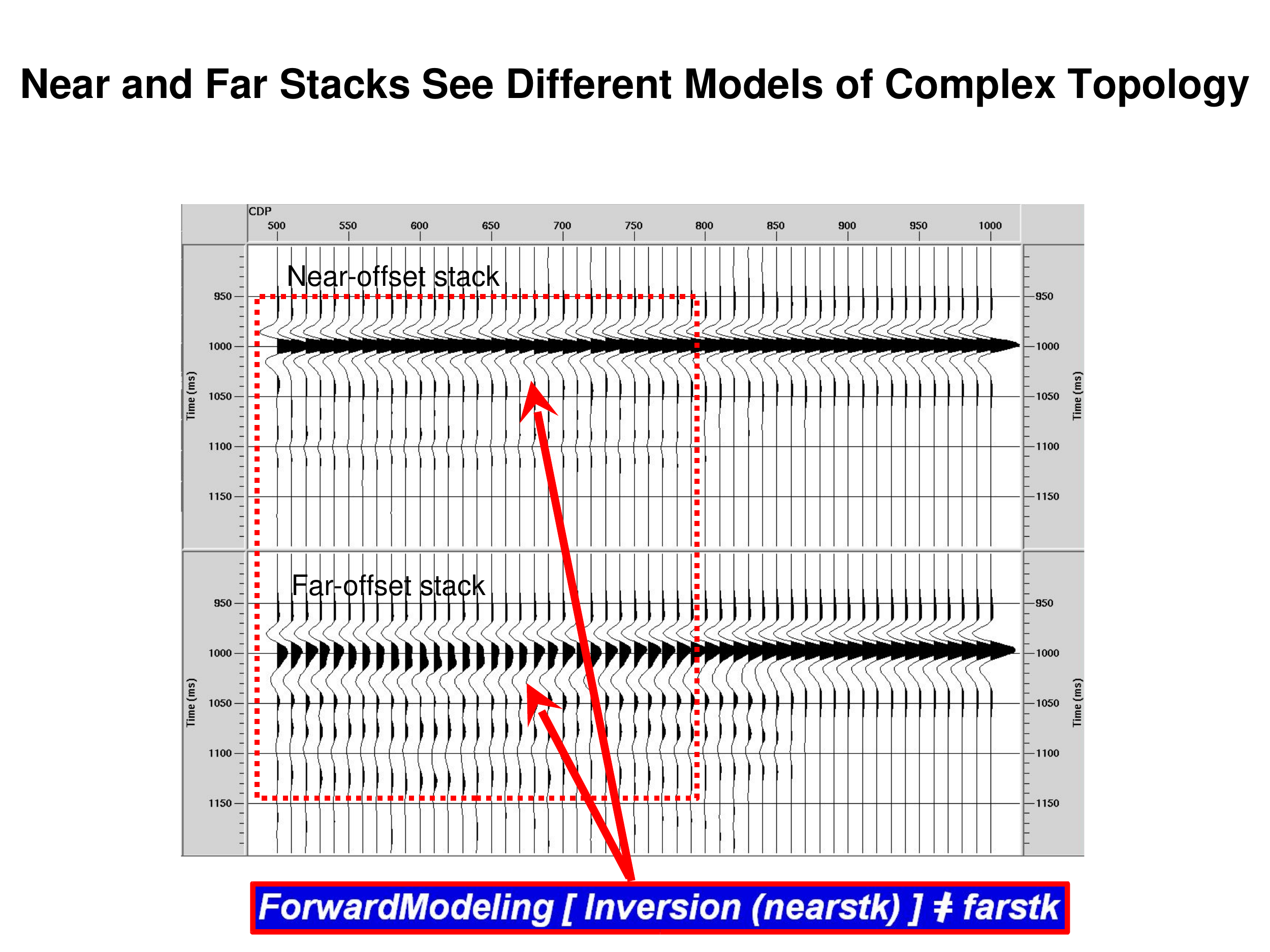}
\caption{Slide 10}
\end{figure}

\clearpage

\begin{figure}
\centering
  \includegraphics[width=5.0in]{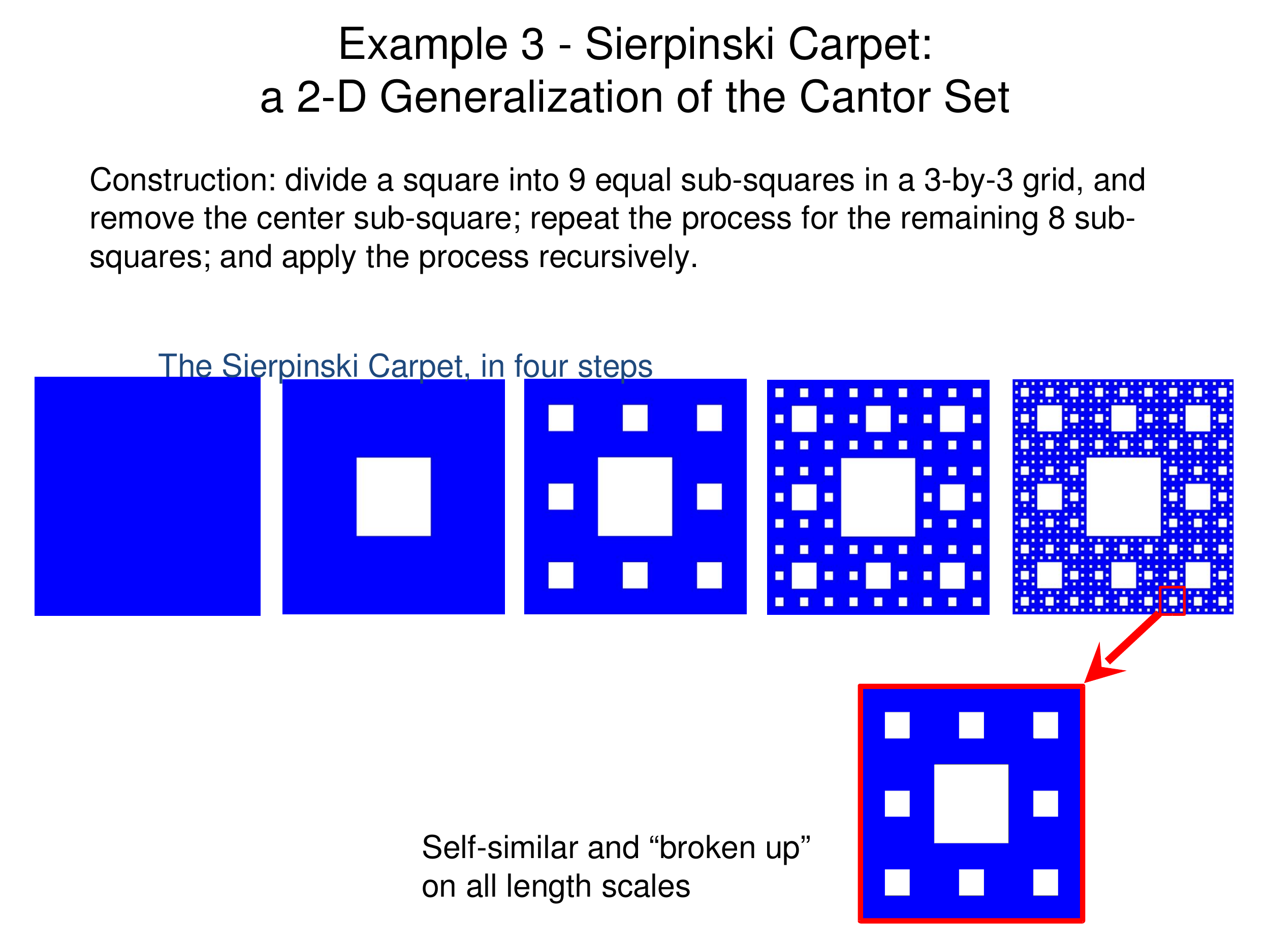}
\caption{Slide 11}
\end{figure}

\begin{figure}
\centering
  \includegraphics[width=5.0in]{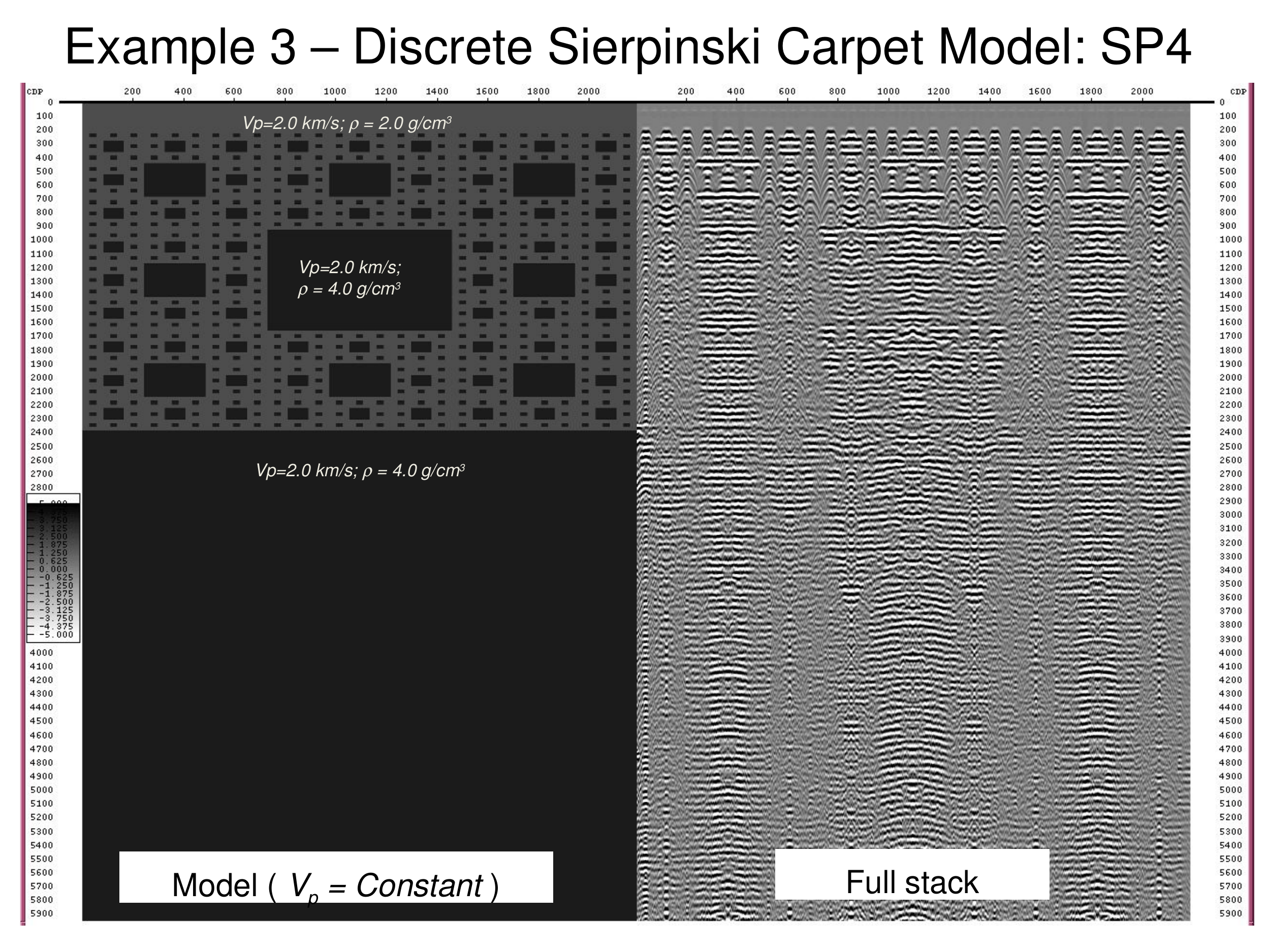}
\caption{Slide 12}
\end{figure}

\begin{figure}
\centering
  \includegraphics[width=5.0in]{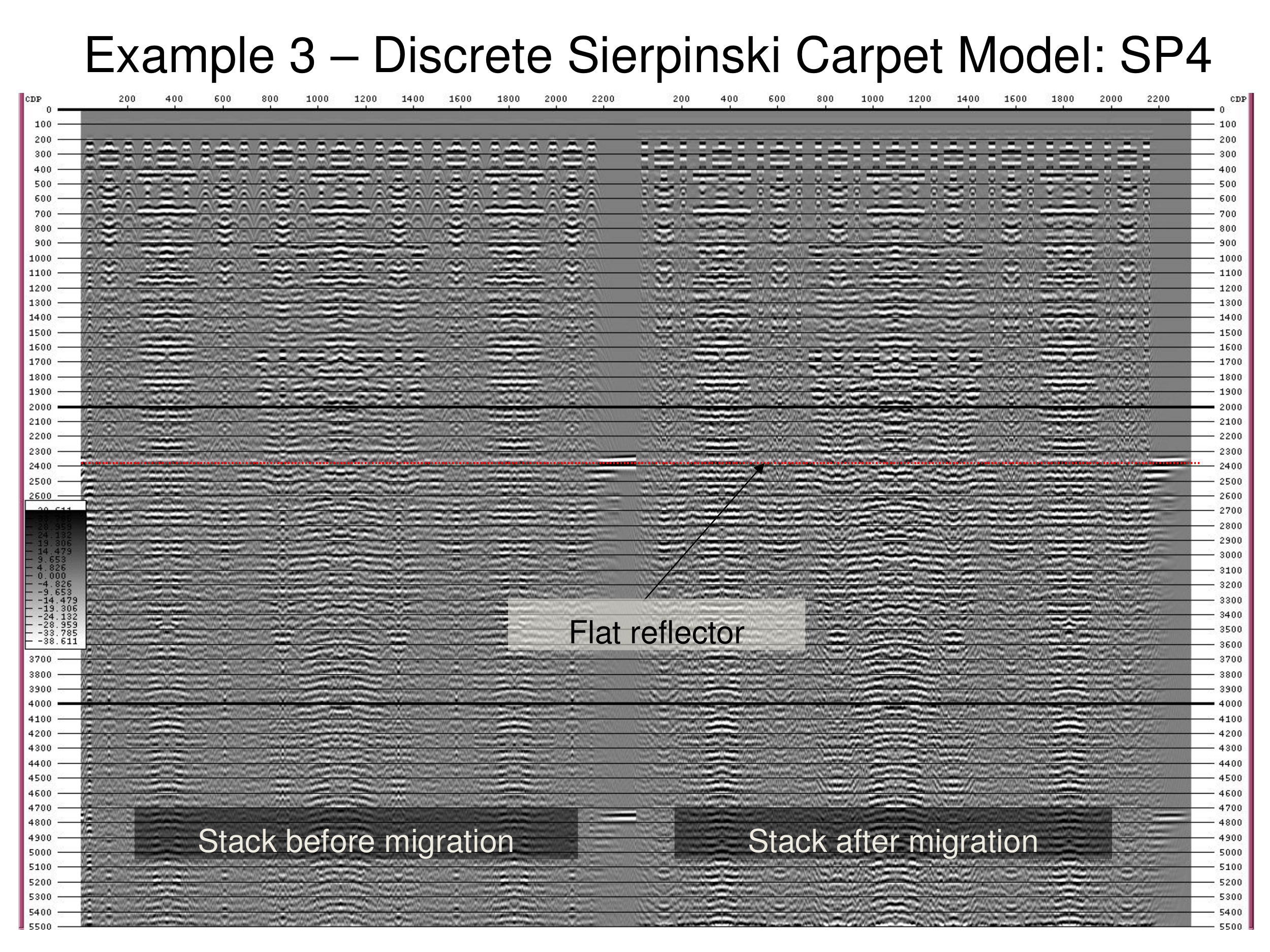}
\caption{Slide 13}
\end{figure}

\begin{figure}
\centering
  \includegraphics[width=5.0in]{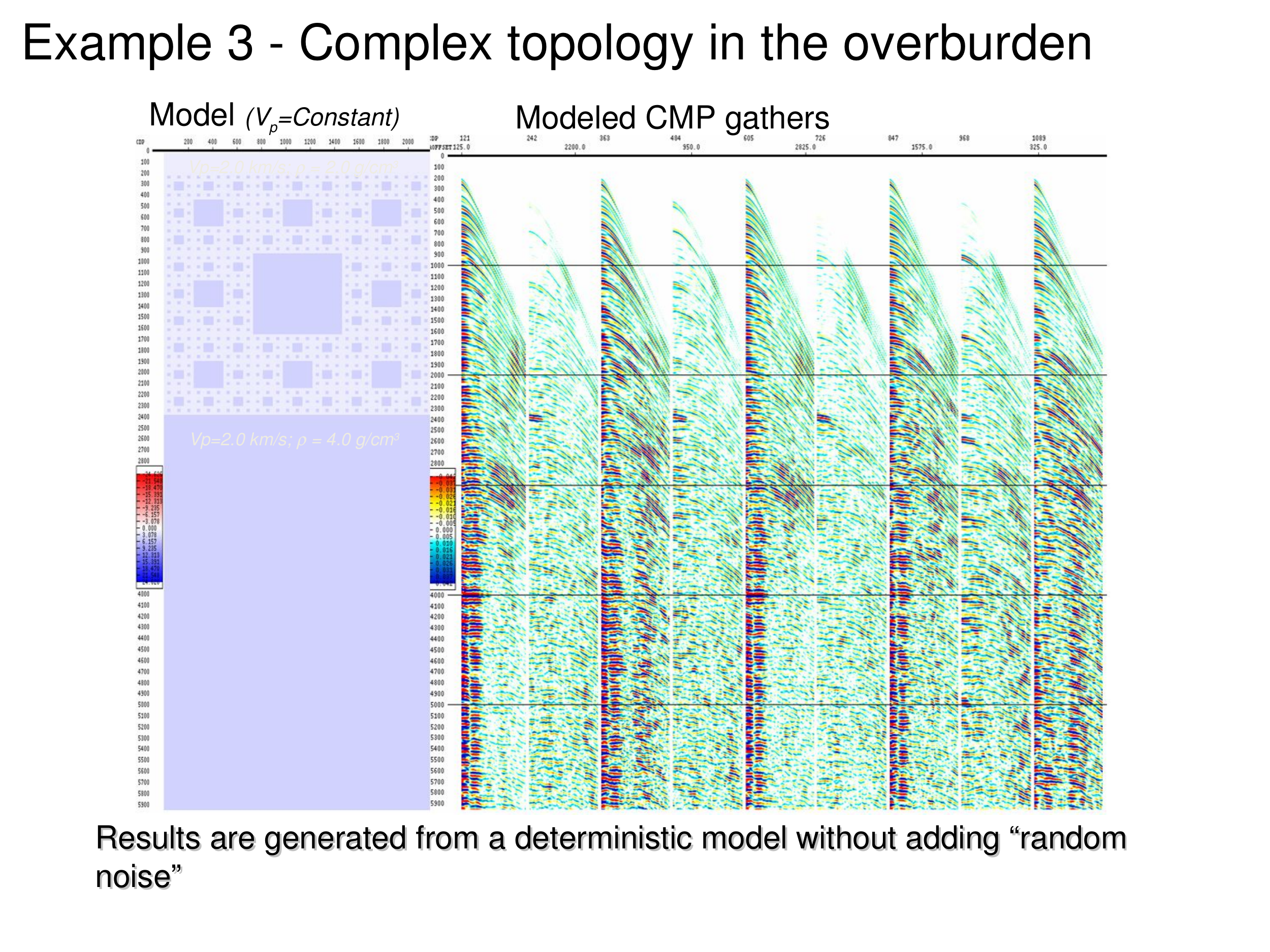}
\caption{Slide 14}
\end{figure}

\begin{figure}
\centering
  \includegraphics[width=5.0in]{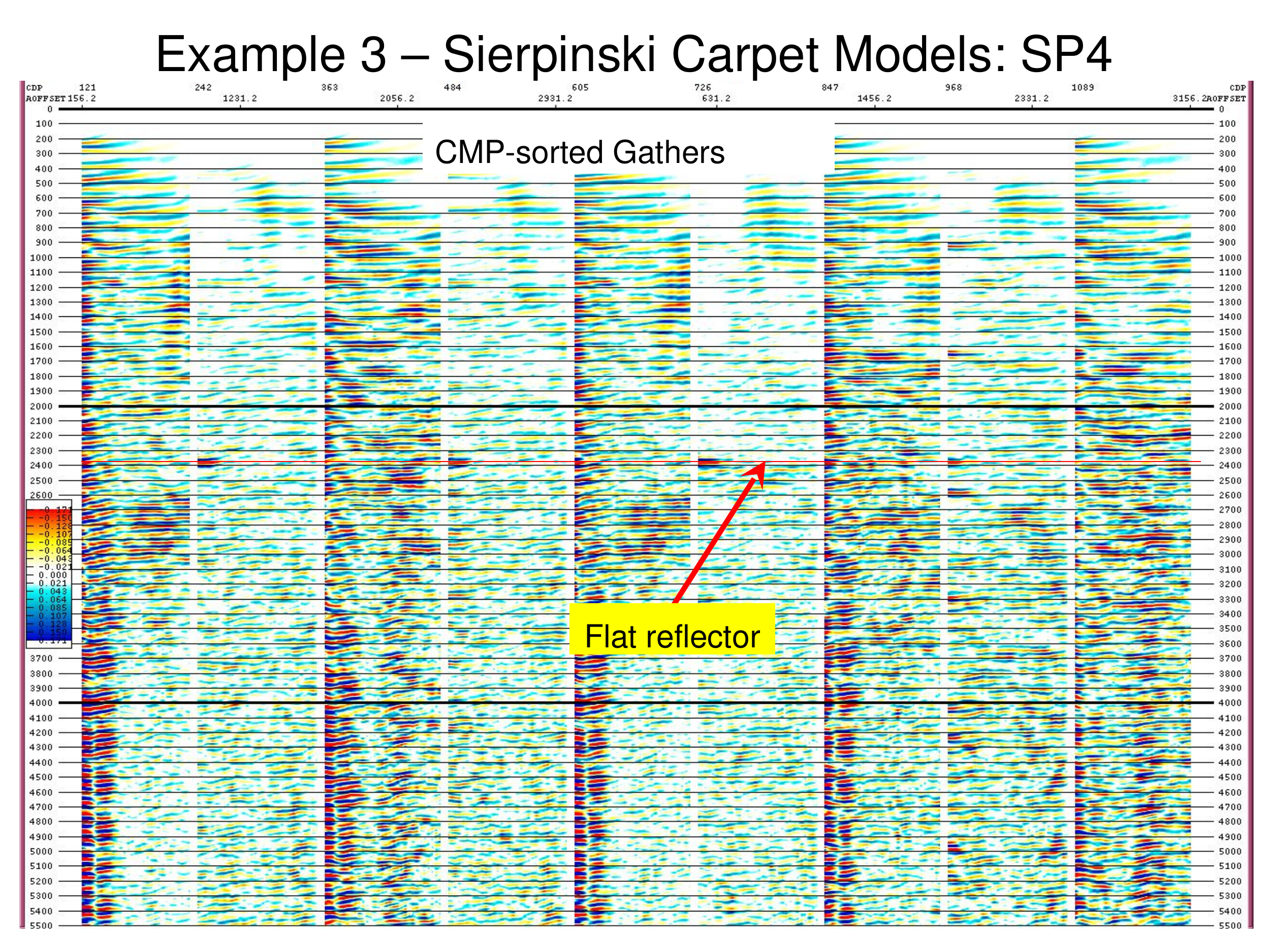}
\caption{Slide 15}
\end{figure}

\begin{figure}
\centering
  \includegraphics[width=5.0in]{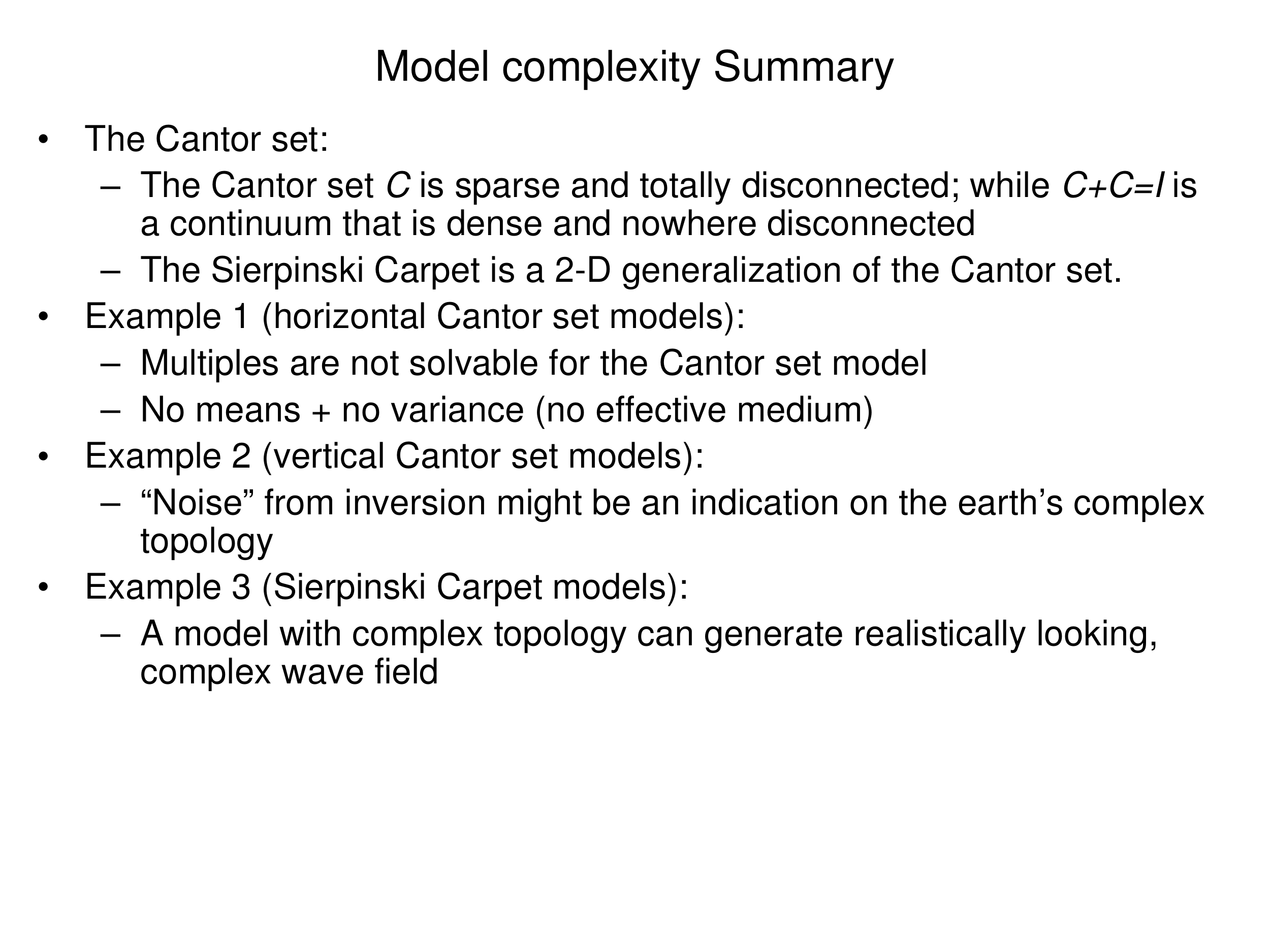}
\caption{Slide 16}
\end{figure}

\begin{figure}
\centering
  \includegraphics[width=5.0in]{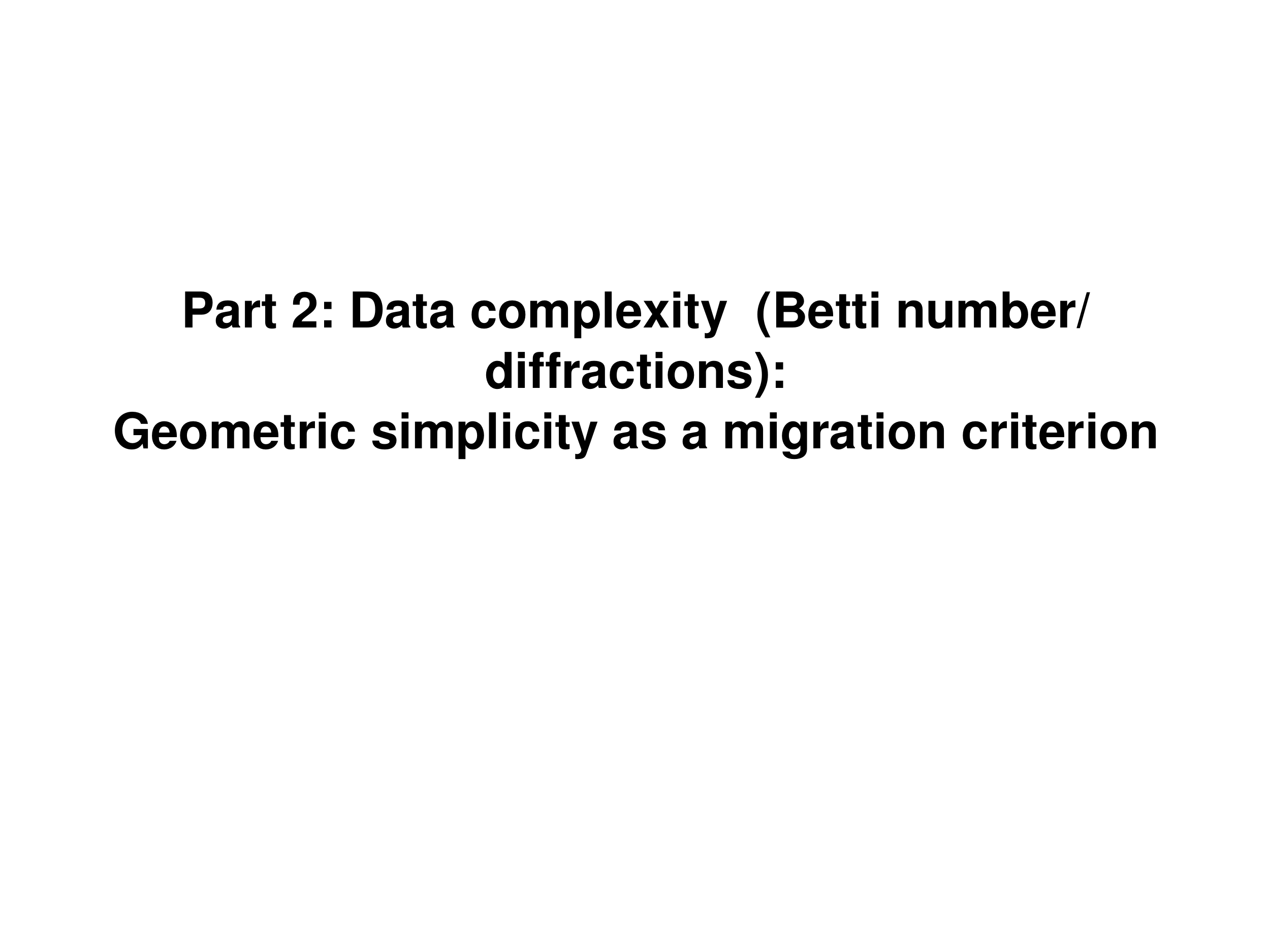}
\caption{Slide 17}
\end{figure}

\begin{figure}
\centering
  \includegraphics[width=5.0in]{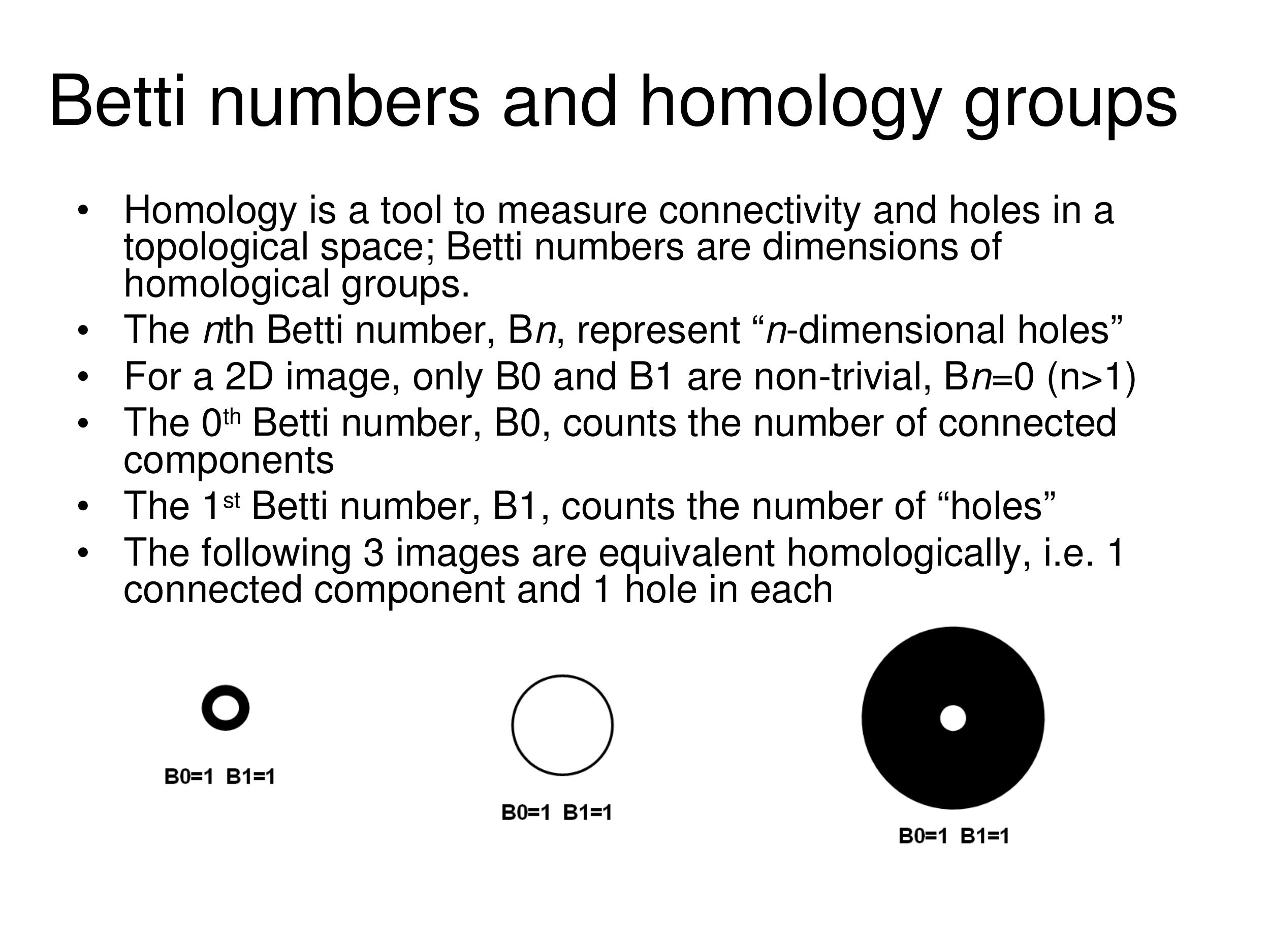}
\caption{Slide 18}
\end{figure}

\begin{figure}
\centering
  \includegraphics[width=5.0in]{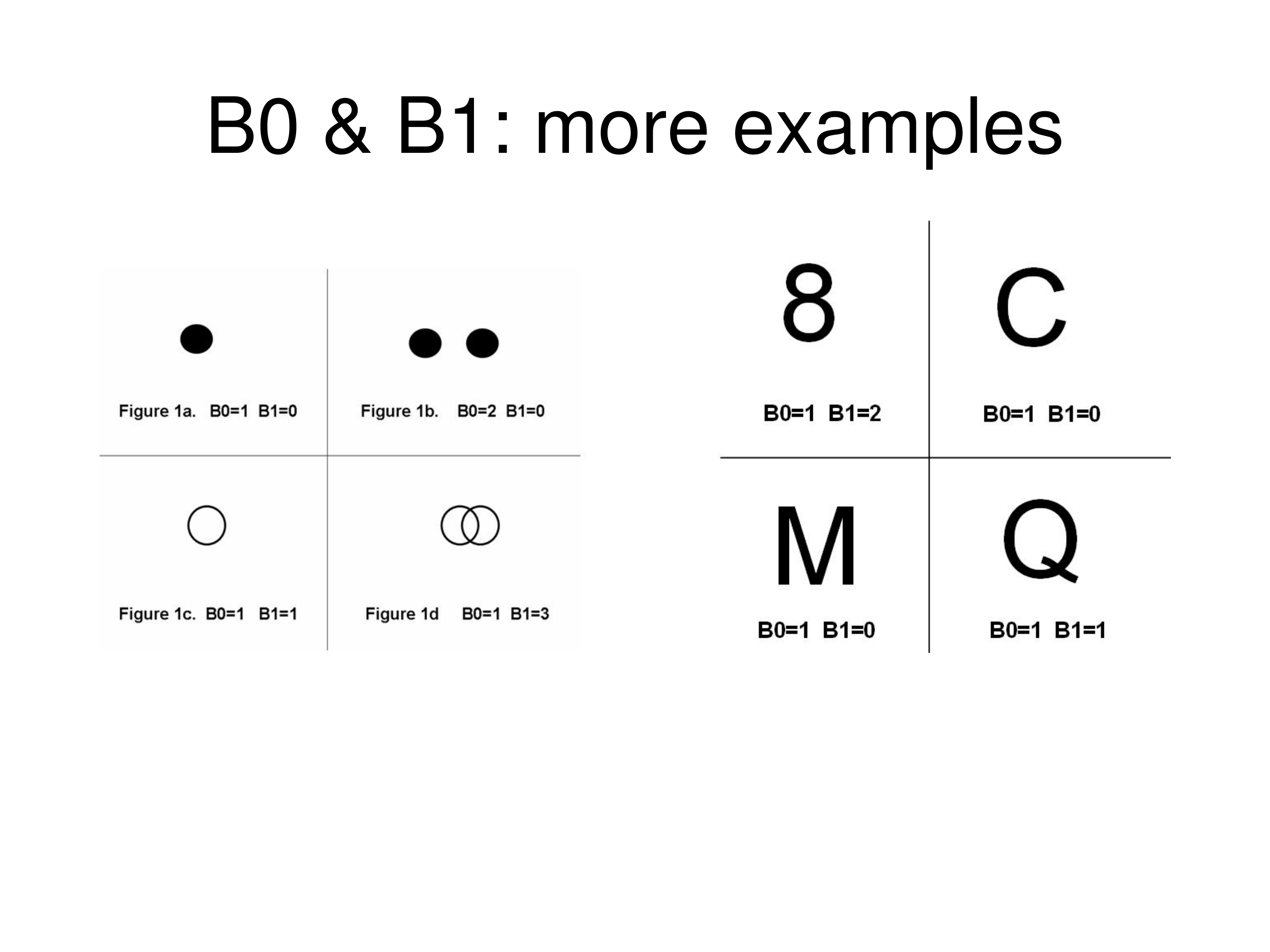}
\caption{Slide 19}
\end{figure}

\begin{figure}
\centering
  \includegraphics[width=5.0in]{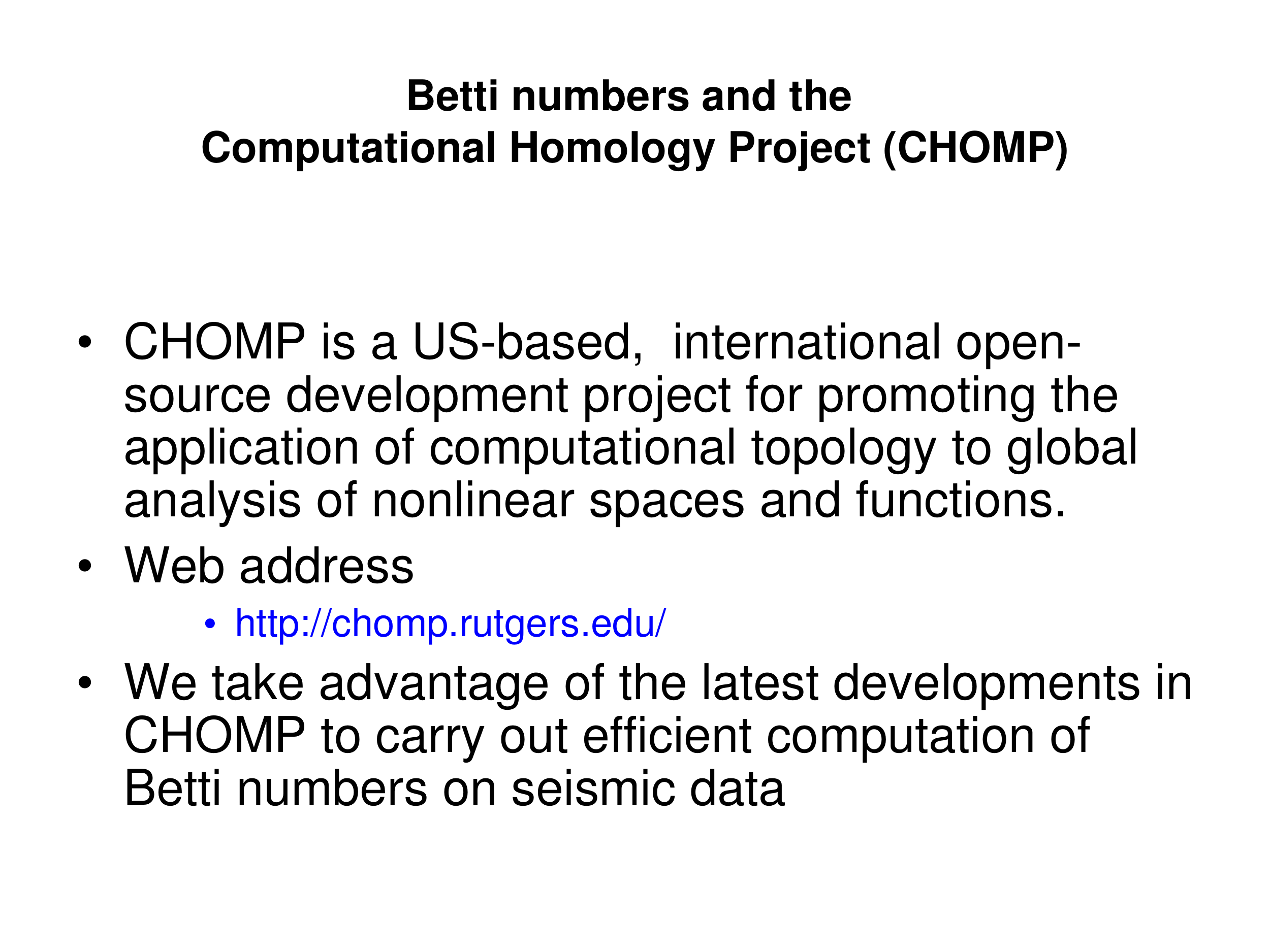}
\caption{Slide 20}
\end{figure}

\clearpage

\begin{figure}
\centering
  \includegraphics[width=5.0in]{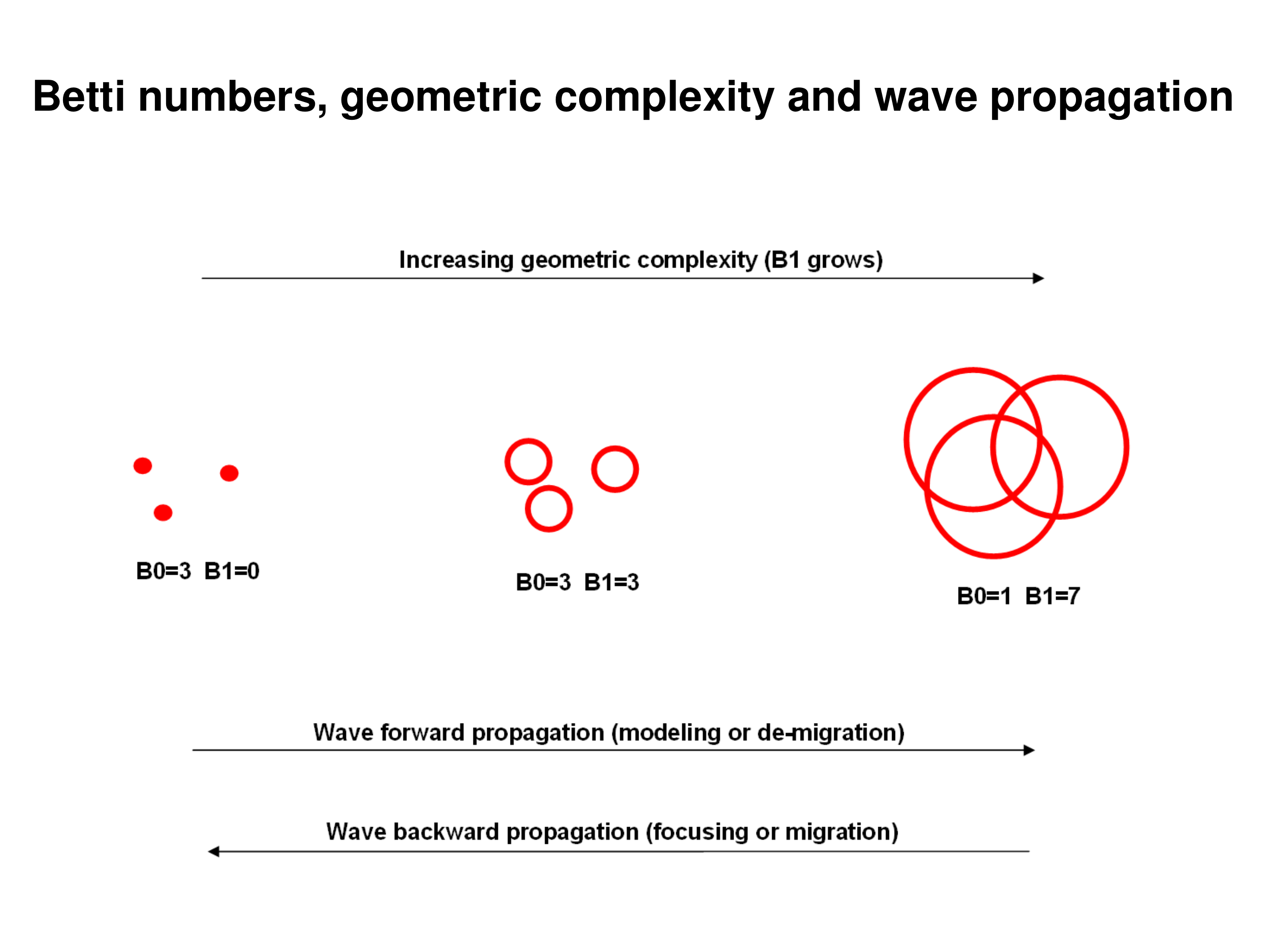}
\caption{Slide 21}
\end{figure}

\begin{figure}
\centering
  \includegraphics[width=5.0in]{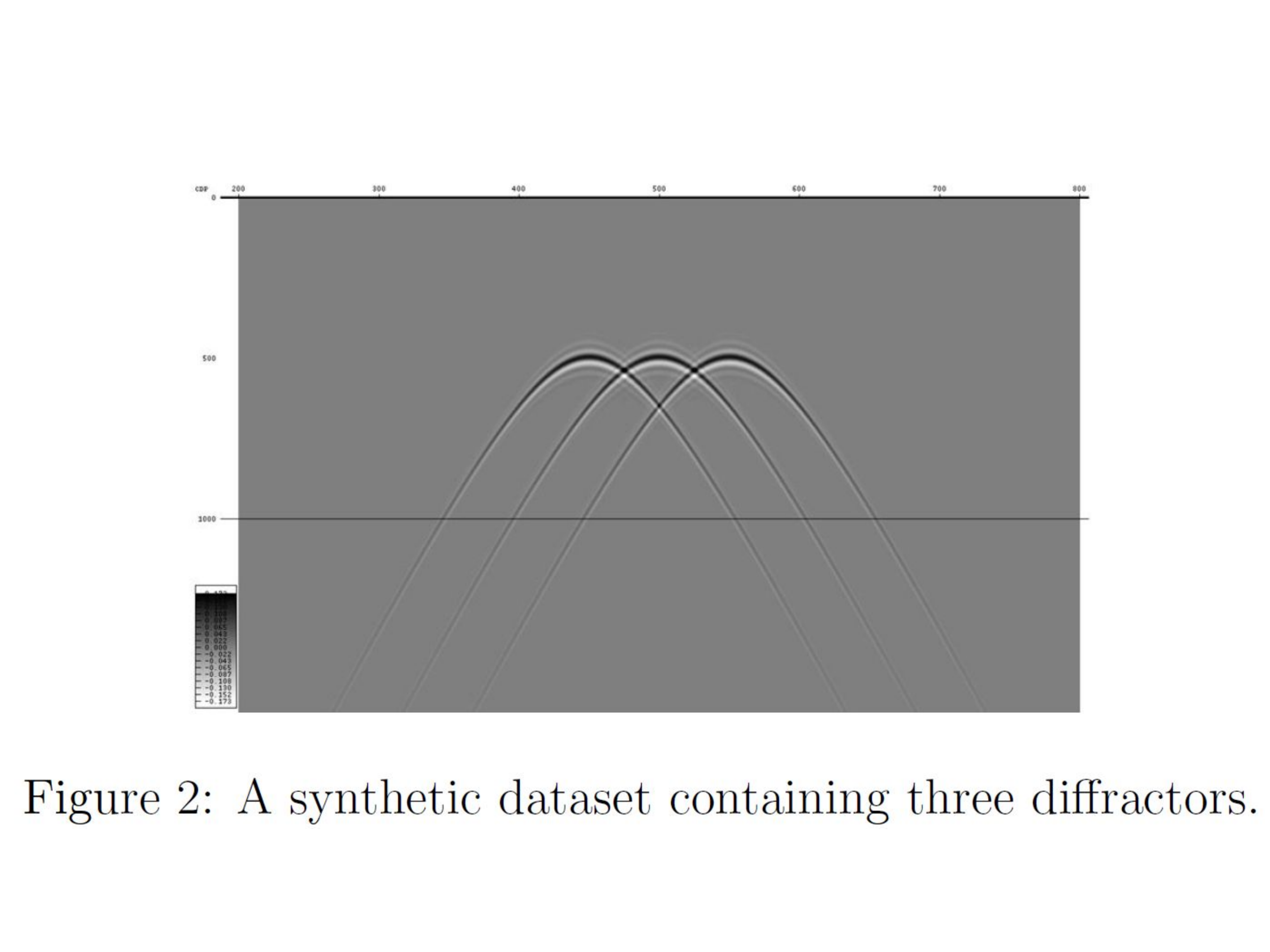}
\caption{Slide 22}
\end{figure}

\begin{figure}
\centering
  \includegraphics[width=5.0in]{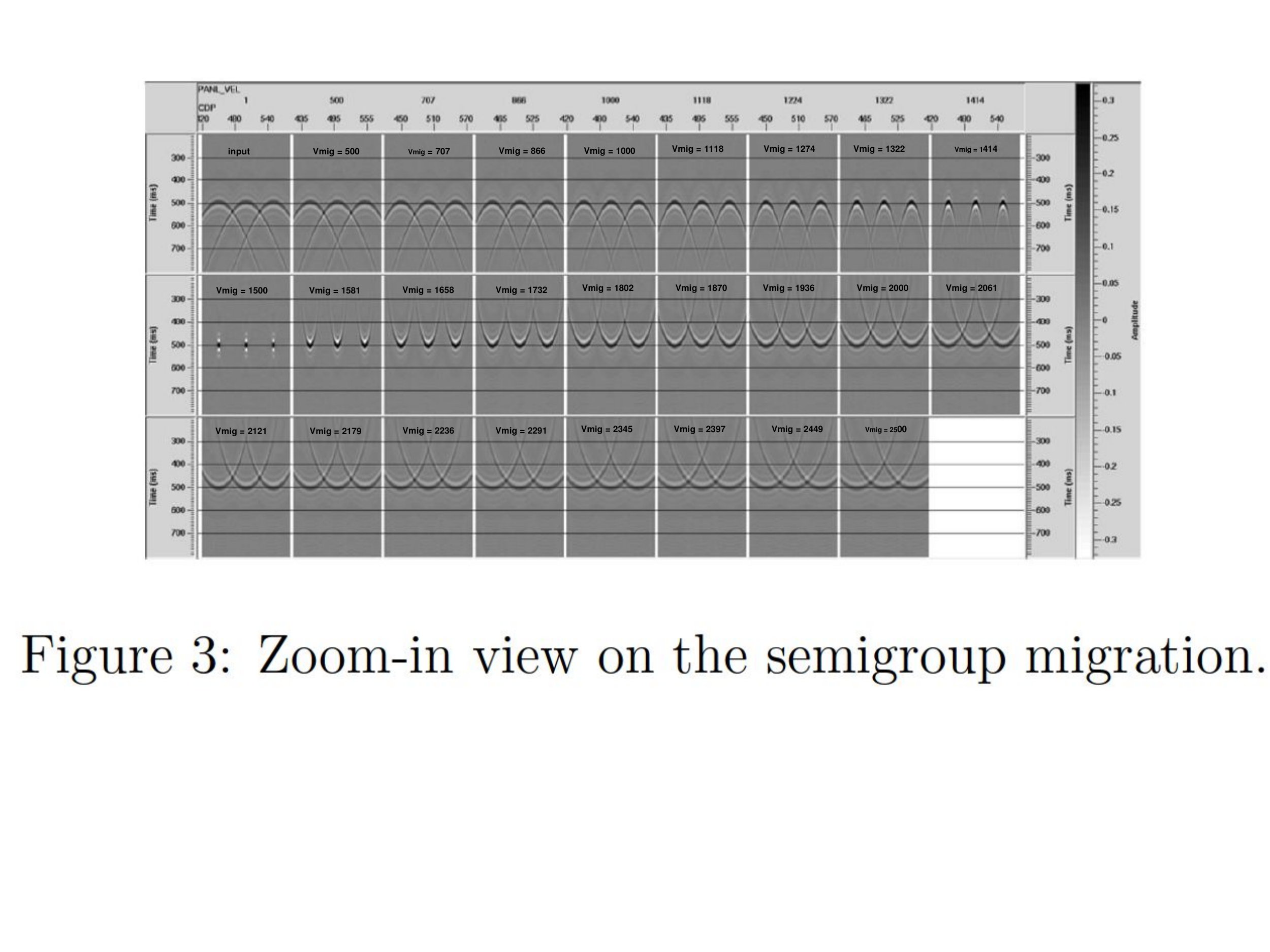}
\caption{Slide 23}
\end{figure}

\begin{figure}
\centering
  \includegraphics[width=5.0in]{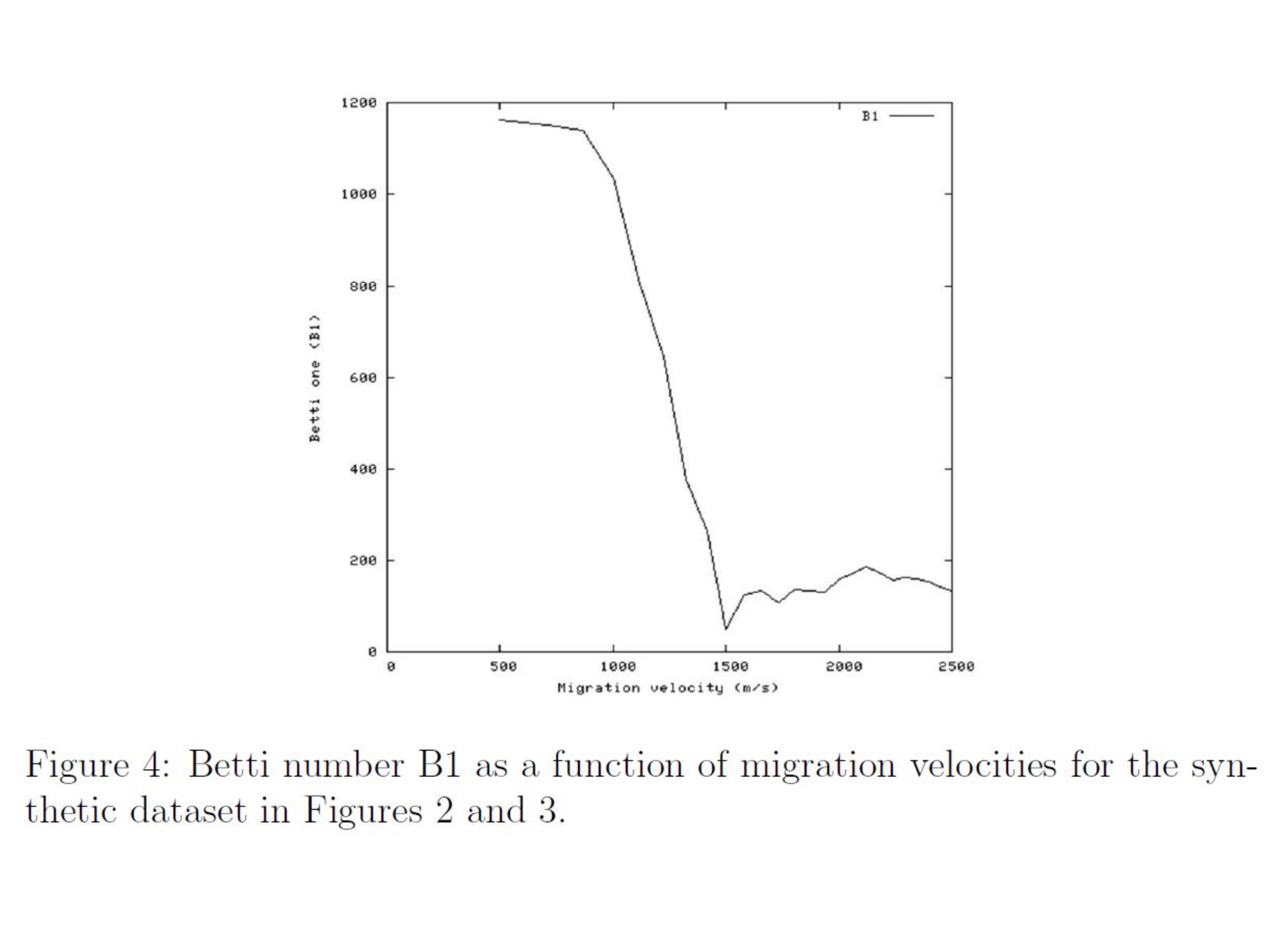}
\caption{Slide 24}
\end{figure}

\begin{figure}
\centering
  \includegraphics[width=5.0in]{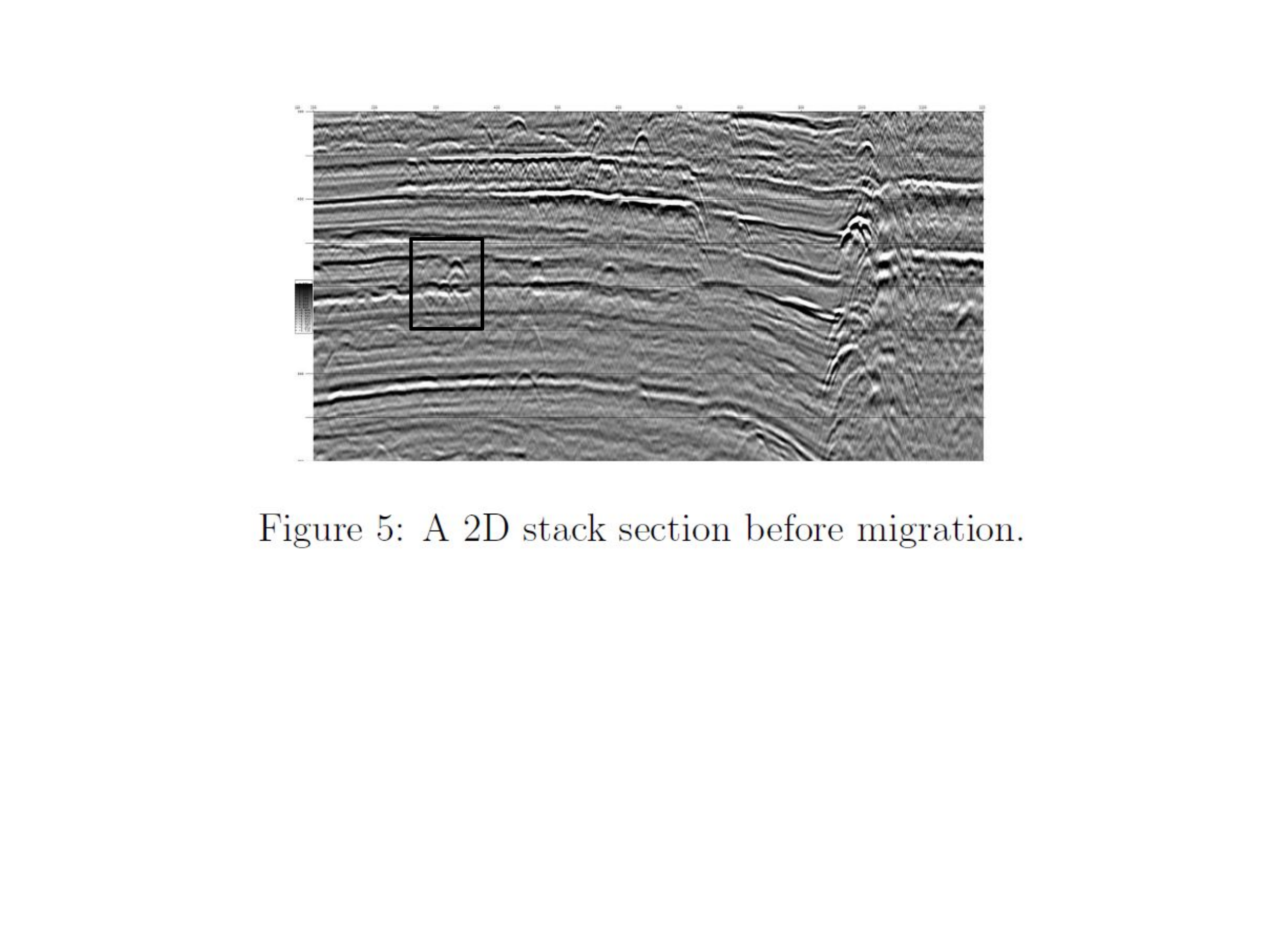}
\caption{Slide 25}
\end{figure}

\begin{figure}
\centering
  \includegraphics[width=5.0in]{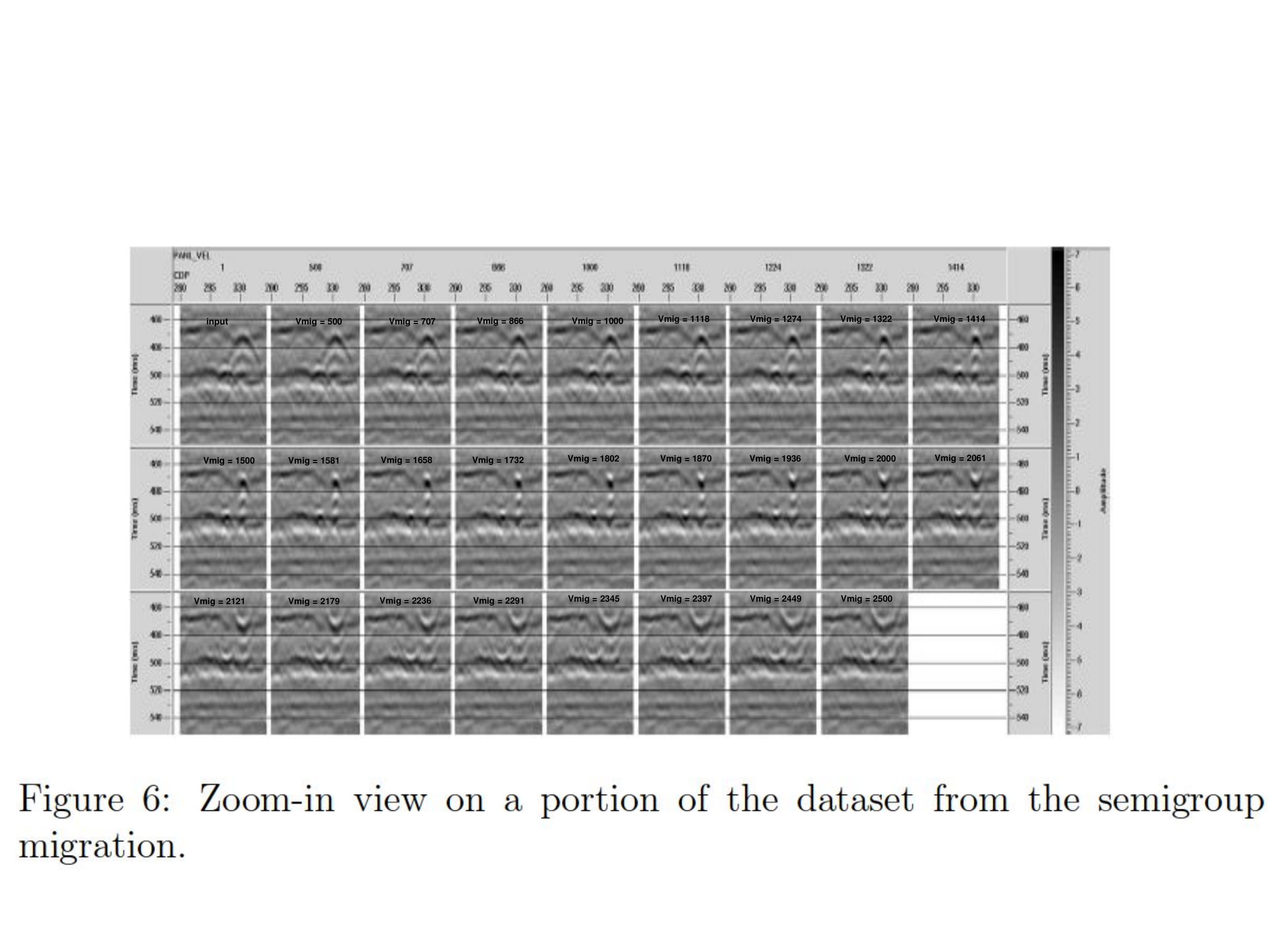}
\caption{Slide 26}
\end{figure}

\begin{figure}
\centering
  \includegraphics[width=5.0in]{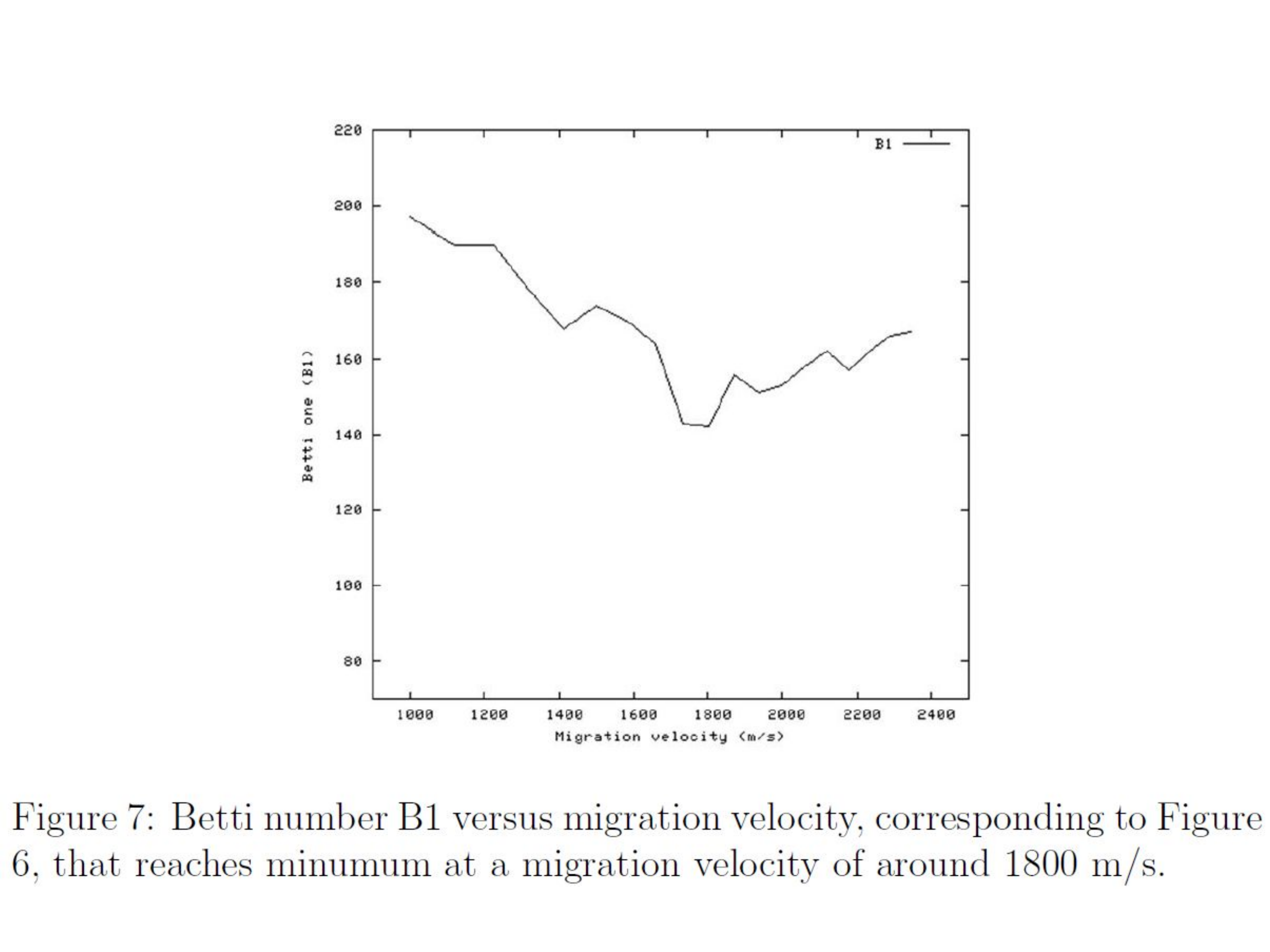}
\caption{Slide 27}
\end{figure}

\begin{figure}
\centering
  \includegraphics[width=5.0in]{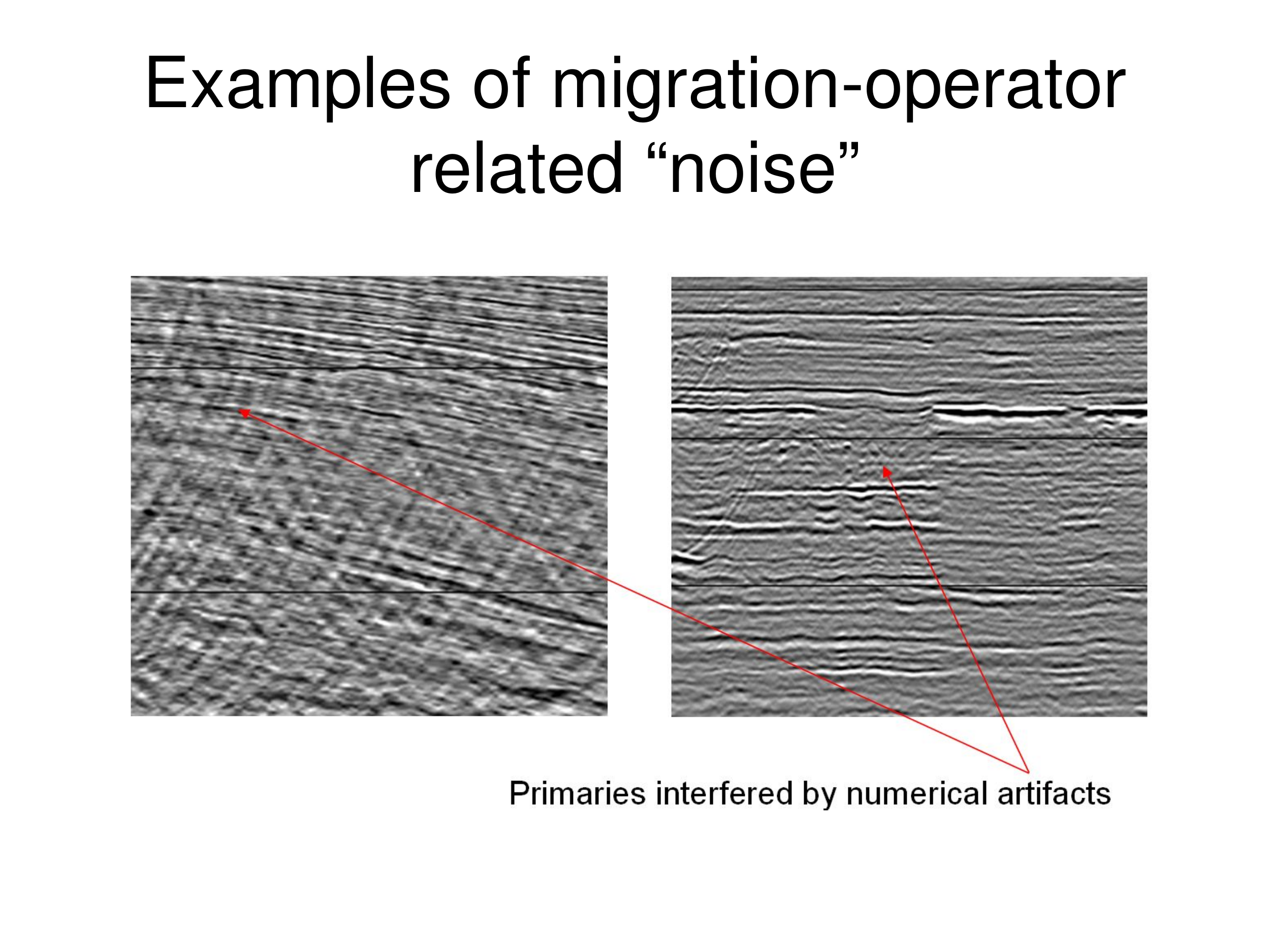}
\caption{Slide 28}
\end{figure}

\begin{figure}
\centering
  \includegraphics[width=5.0in]{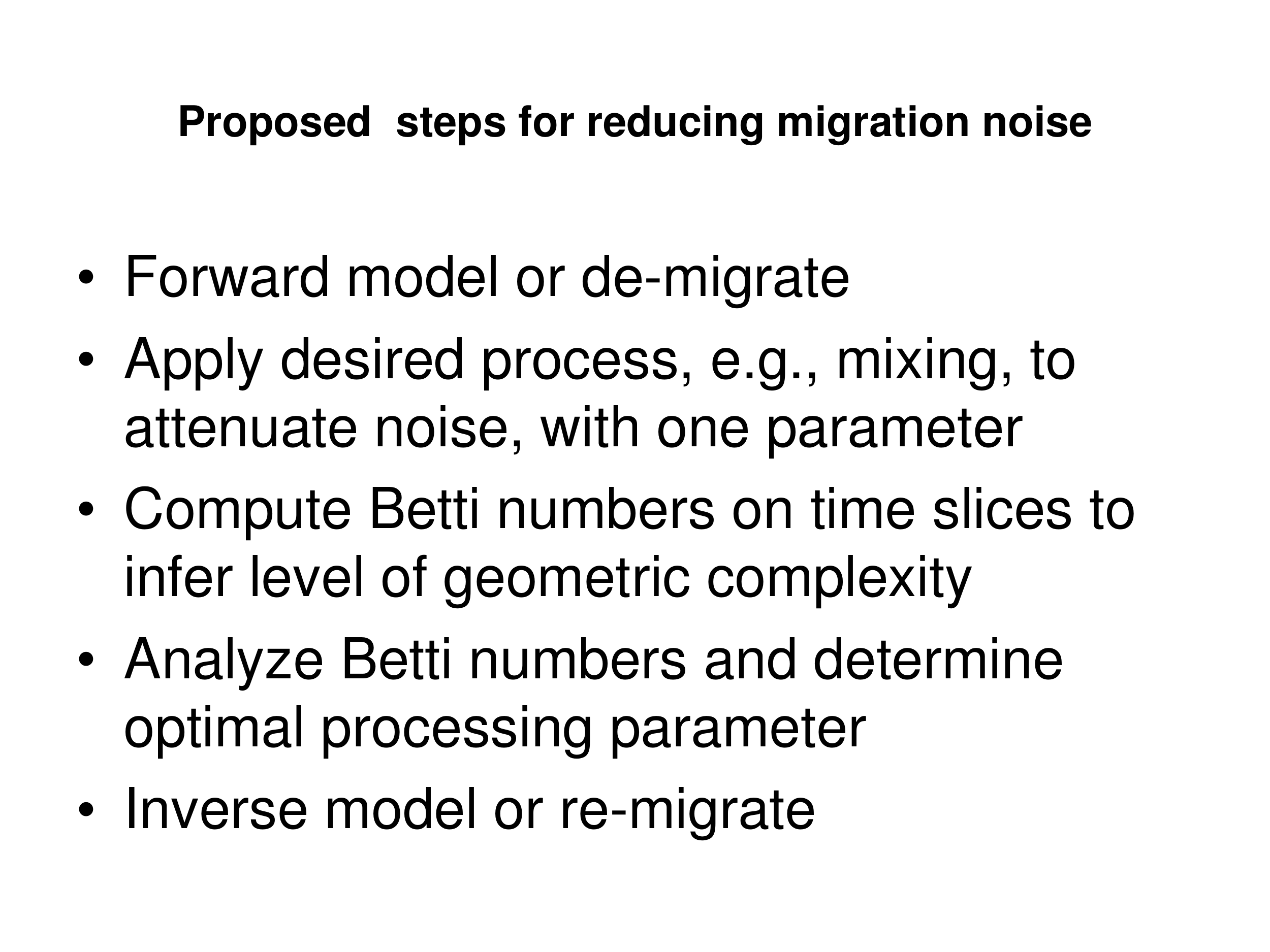}
\caption{Slide 29}
\end{figure}

\begin{figure}
\centering
  \includegraphics[width=5.0in]{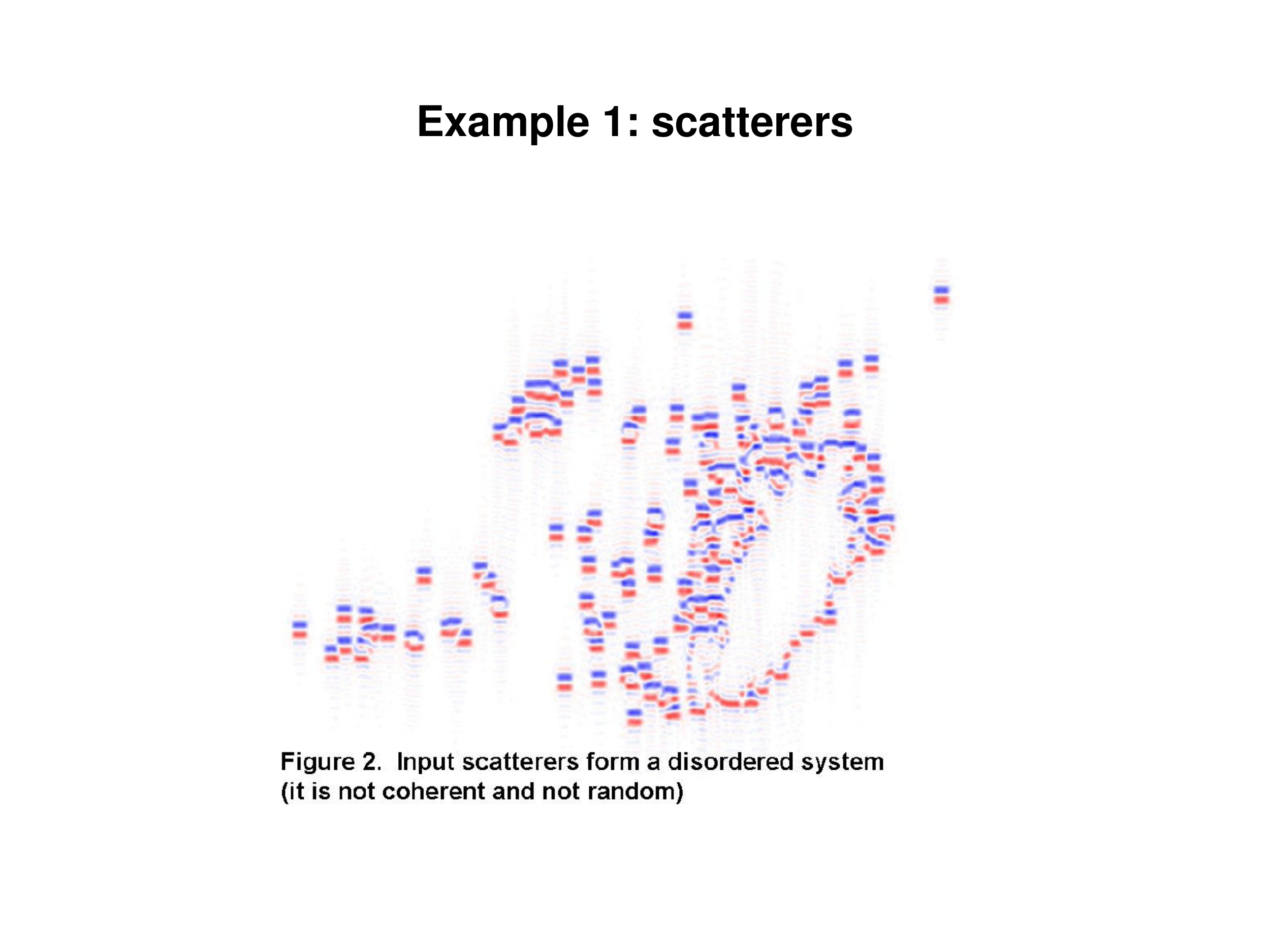}
\caption{Slide 30}
\end{figure}

\clearpage

\begin{figure}
\centering
  \includegraphics[width=5.0in]{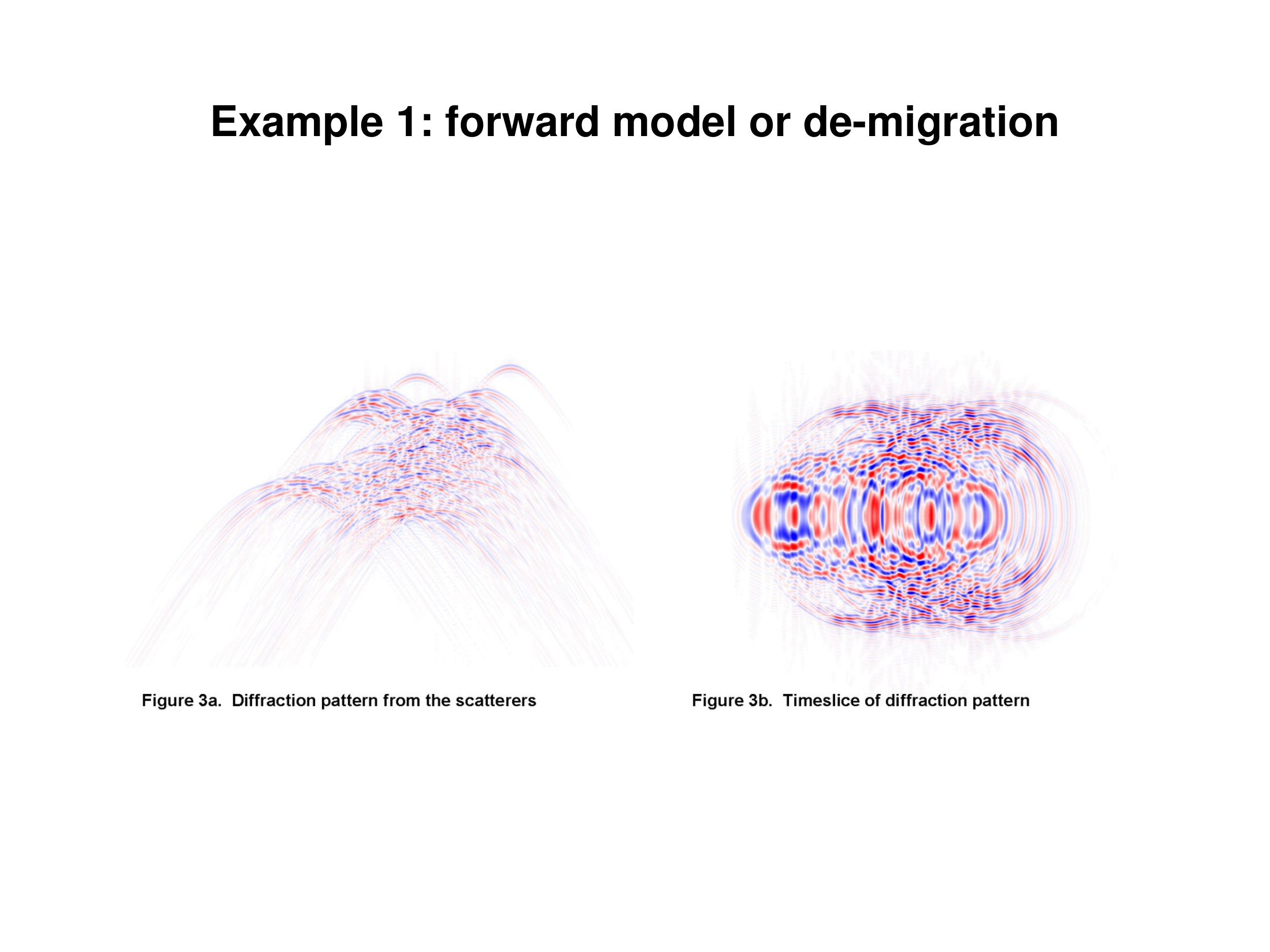}
\caption{Slide 31}
\end{figure}

\begin{figure}
\centering
  \includegraphics[width=5.0in]{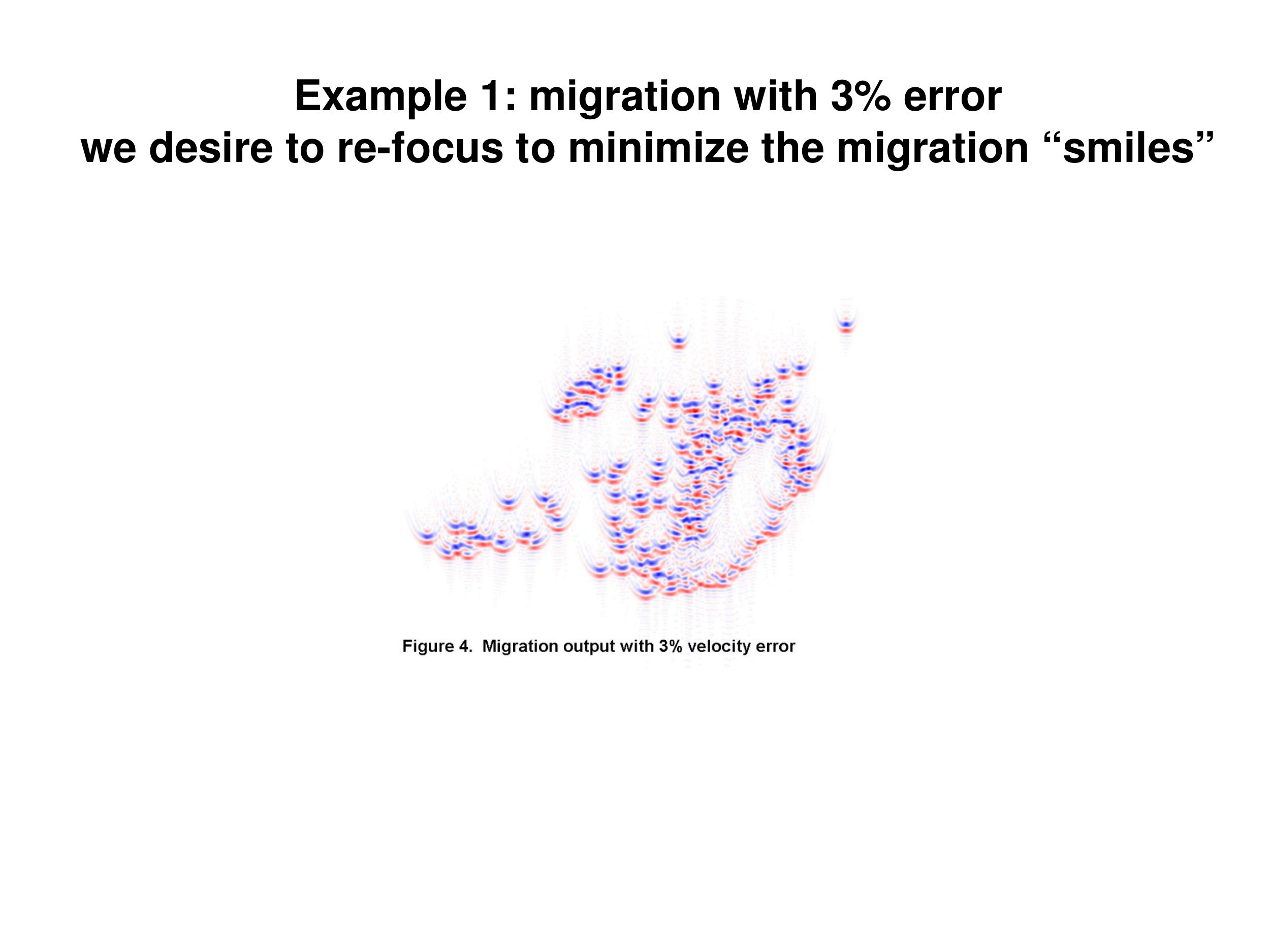}
\caption{Slide 32}
\end{figure}

\begin{figure}
\centering
  \includegraphics[width=5.0in]{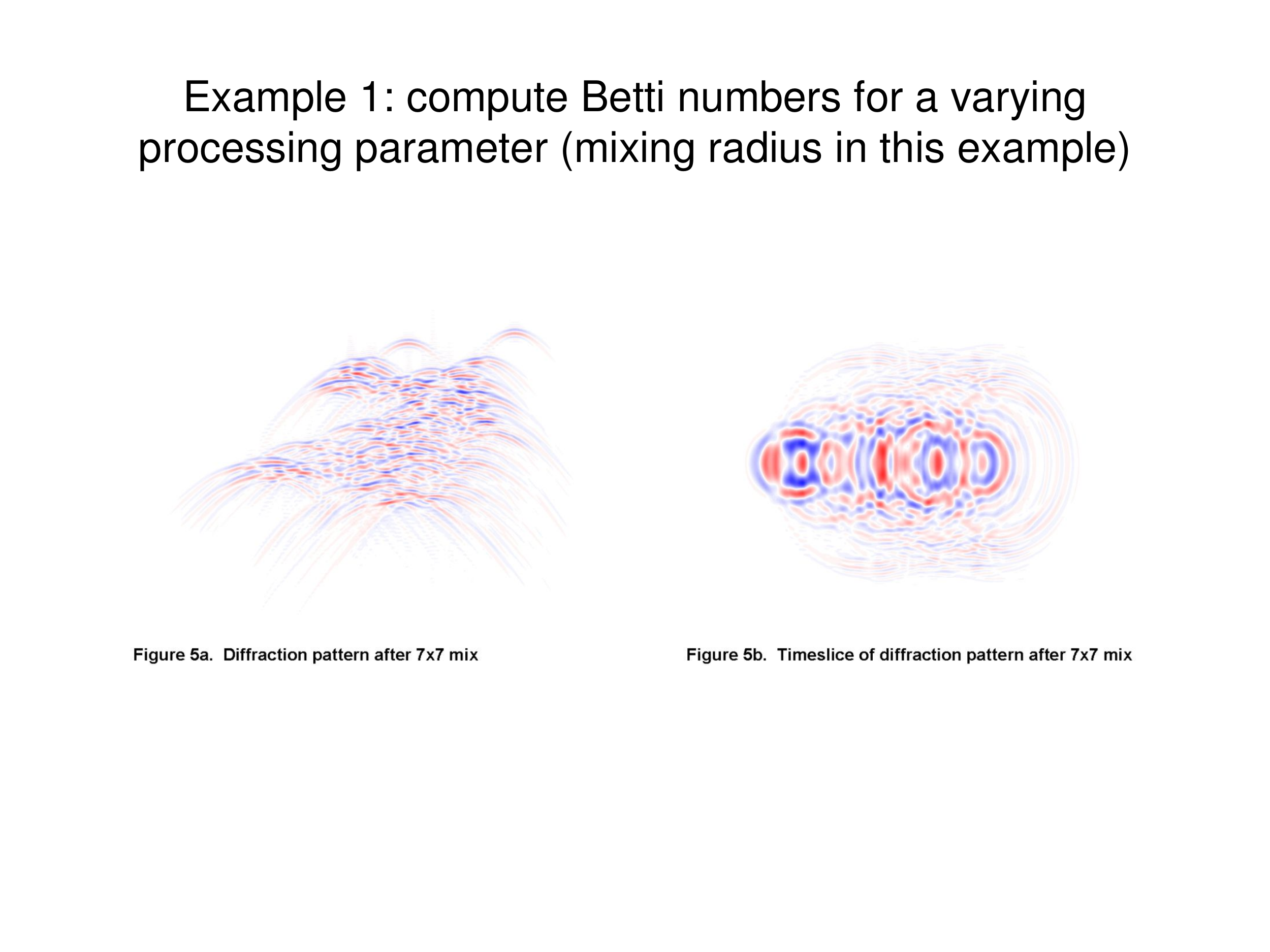}
\caption{Slide 33}
\end{figure}

\begin{figure}
\centering
  \includegraphics[width=5.0in]{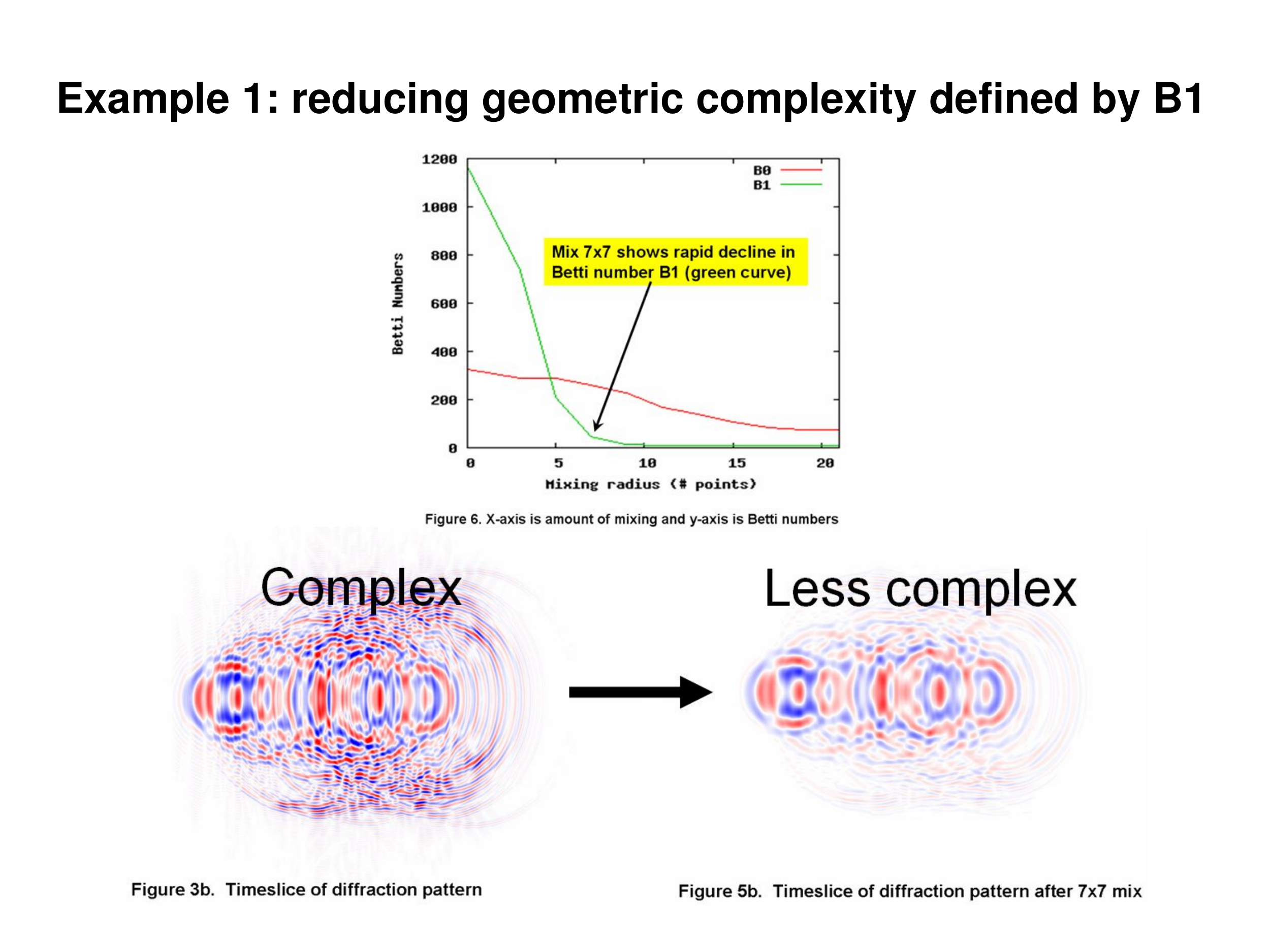}
\caption{Slide 34}
\end{figure}

\begin{figure}
\centering
  \includegraphics[width=5.0in]{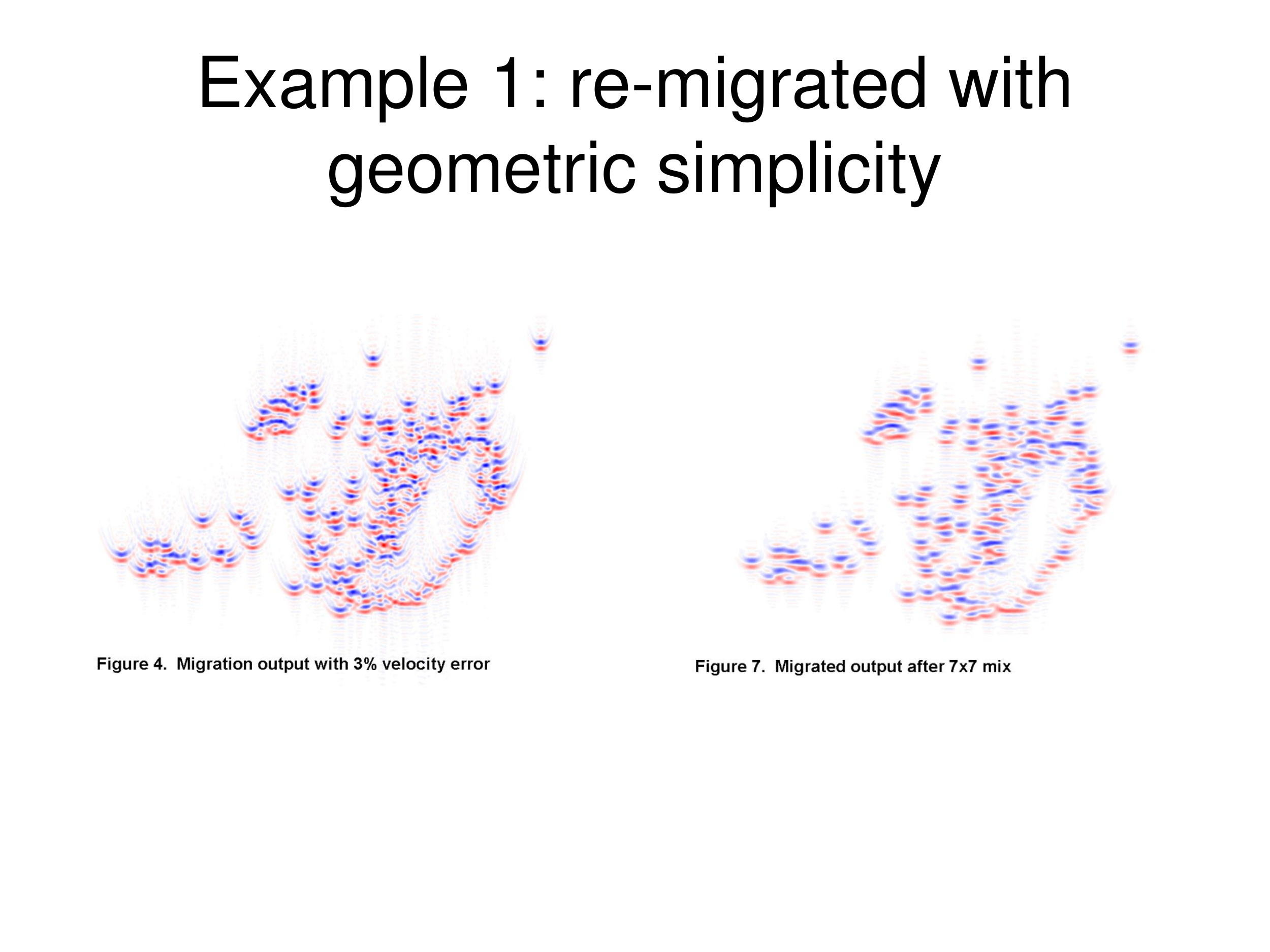}
\caption{Slide 35}
\end{figure}

\begin{figure}
\centering
  \includegraphics[width=5.0in]{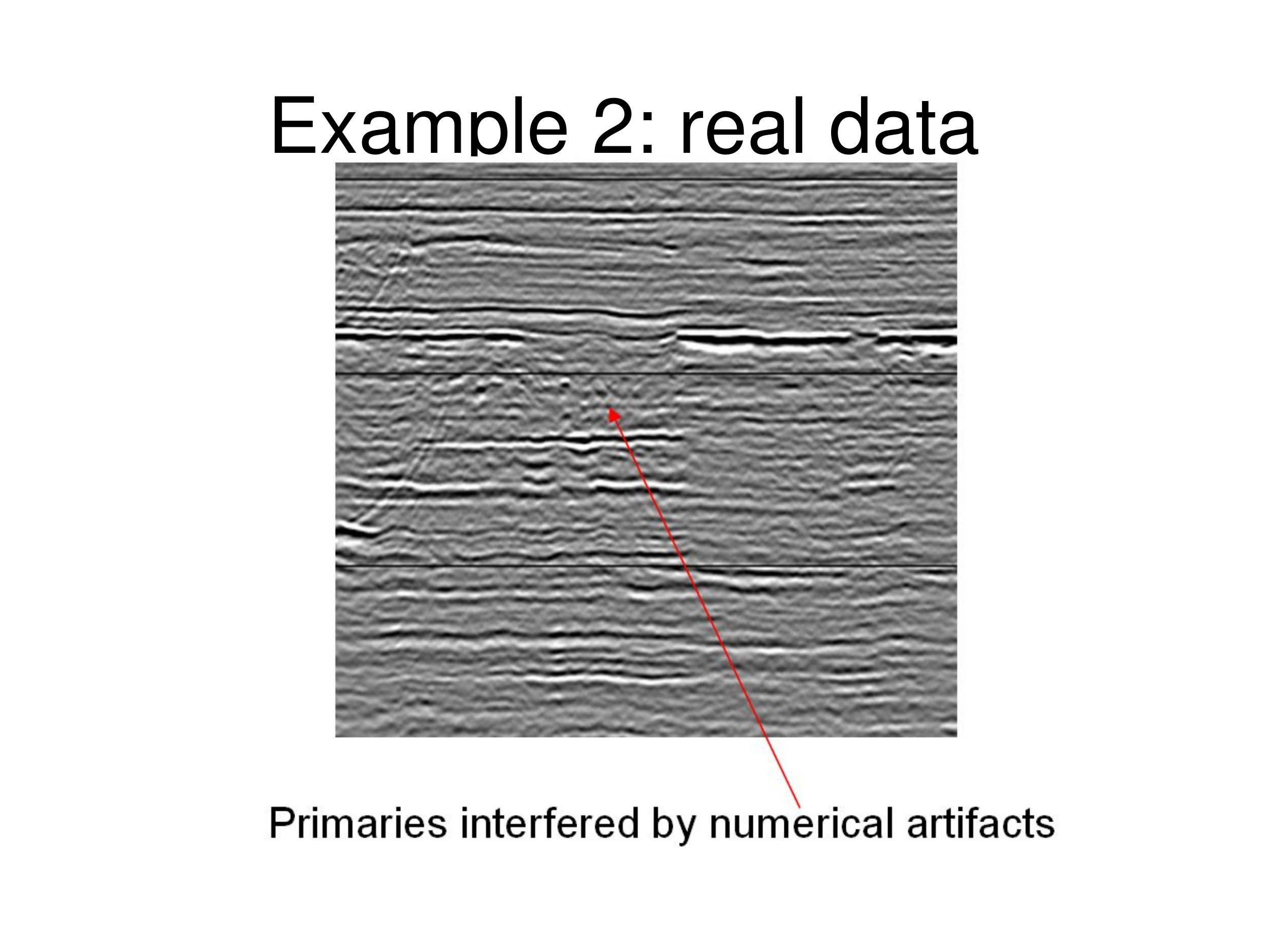}
\caption{Slide 36}
\end{figure}

\begin{figure}
\centering
  \includegraphics[width=5.0in]{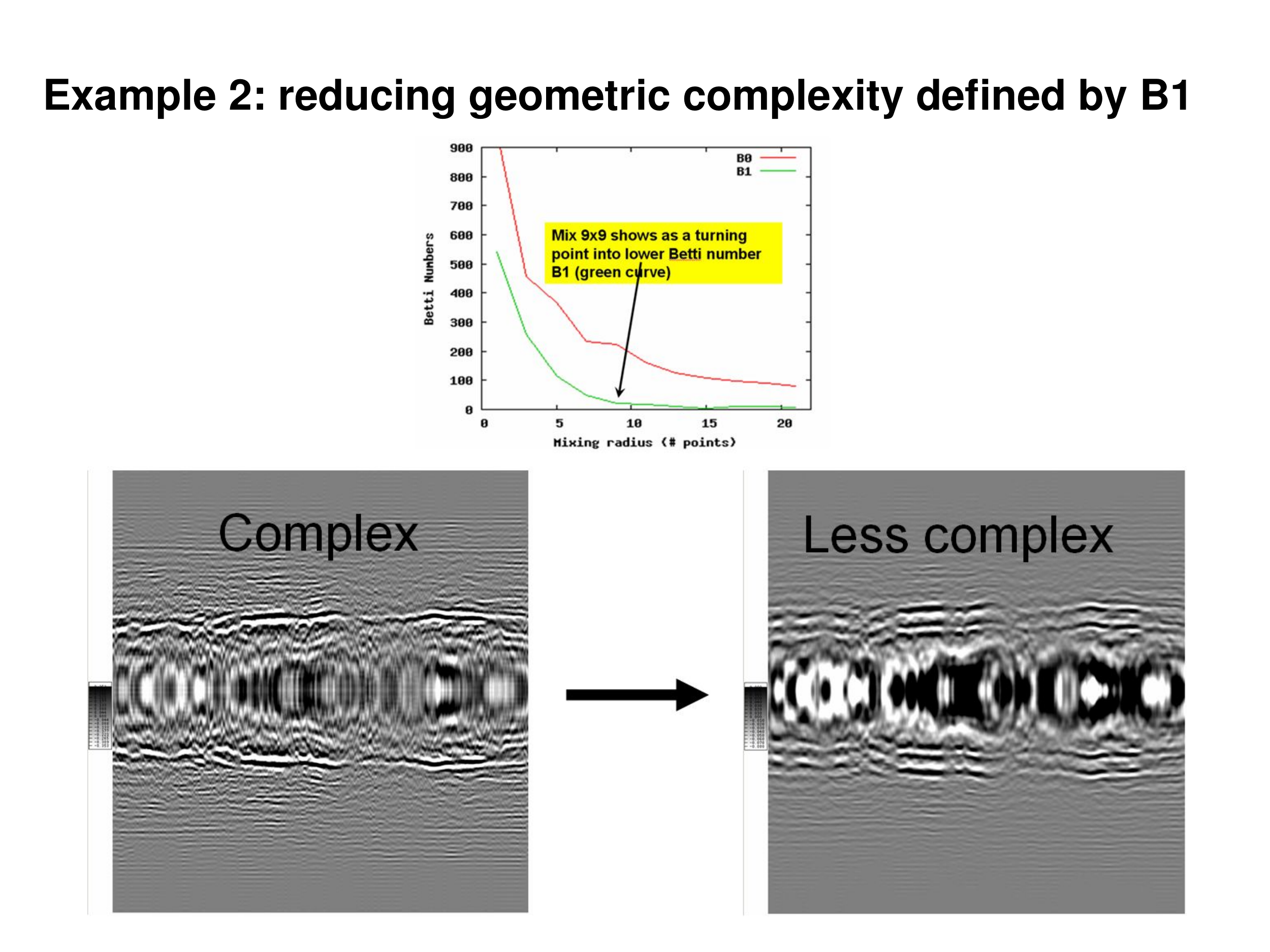}
\caption{Slide 37}
\end{figure}

\begin{figure}
\centering
  \includegraphics[width=5.0in]{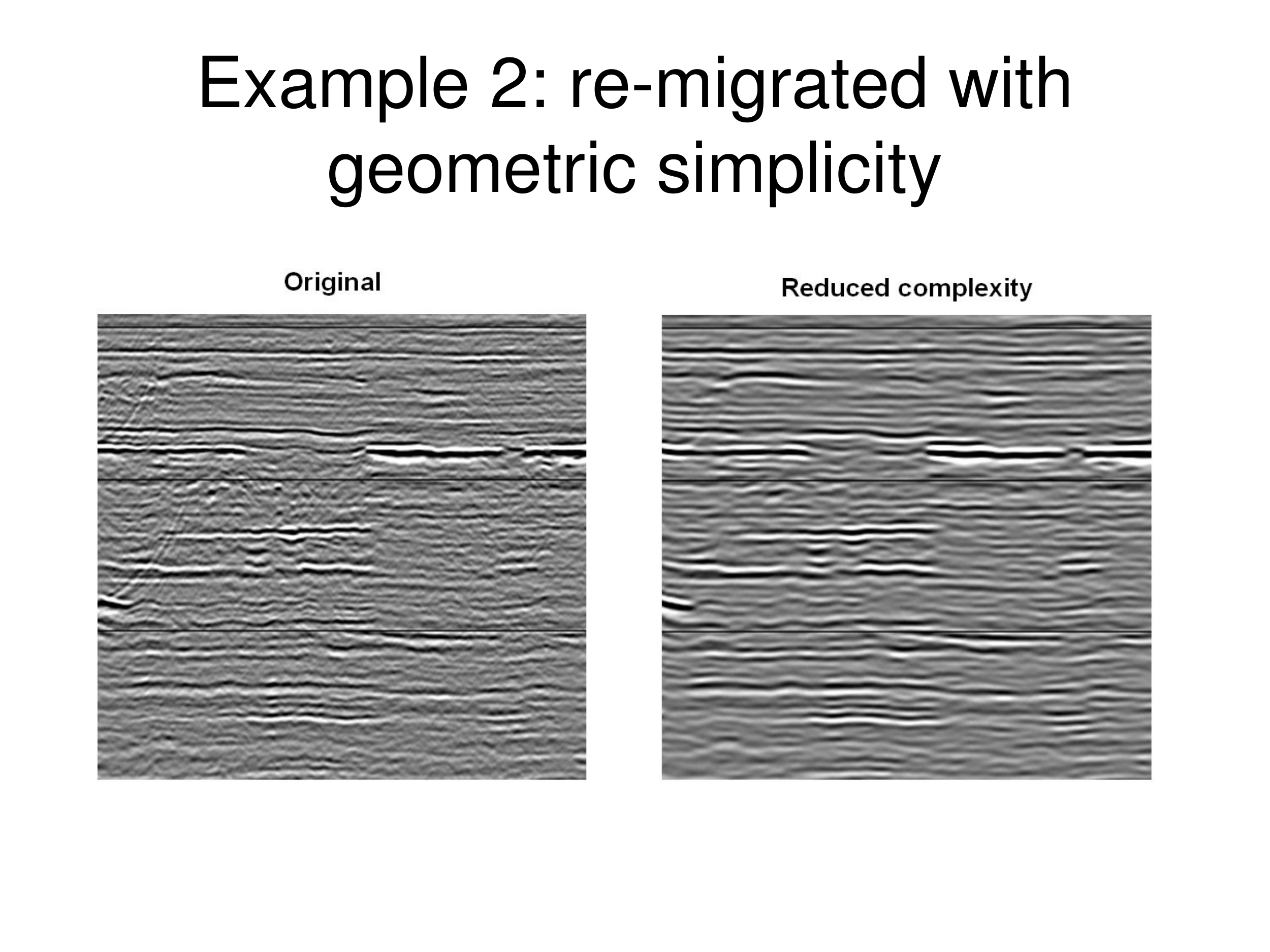}
\caption{Slide 38}
\end{figure}

\begin{figure}
\centering
  \includegraphics[width=5.0in]{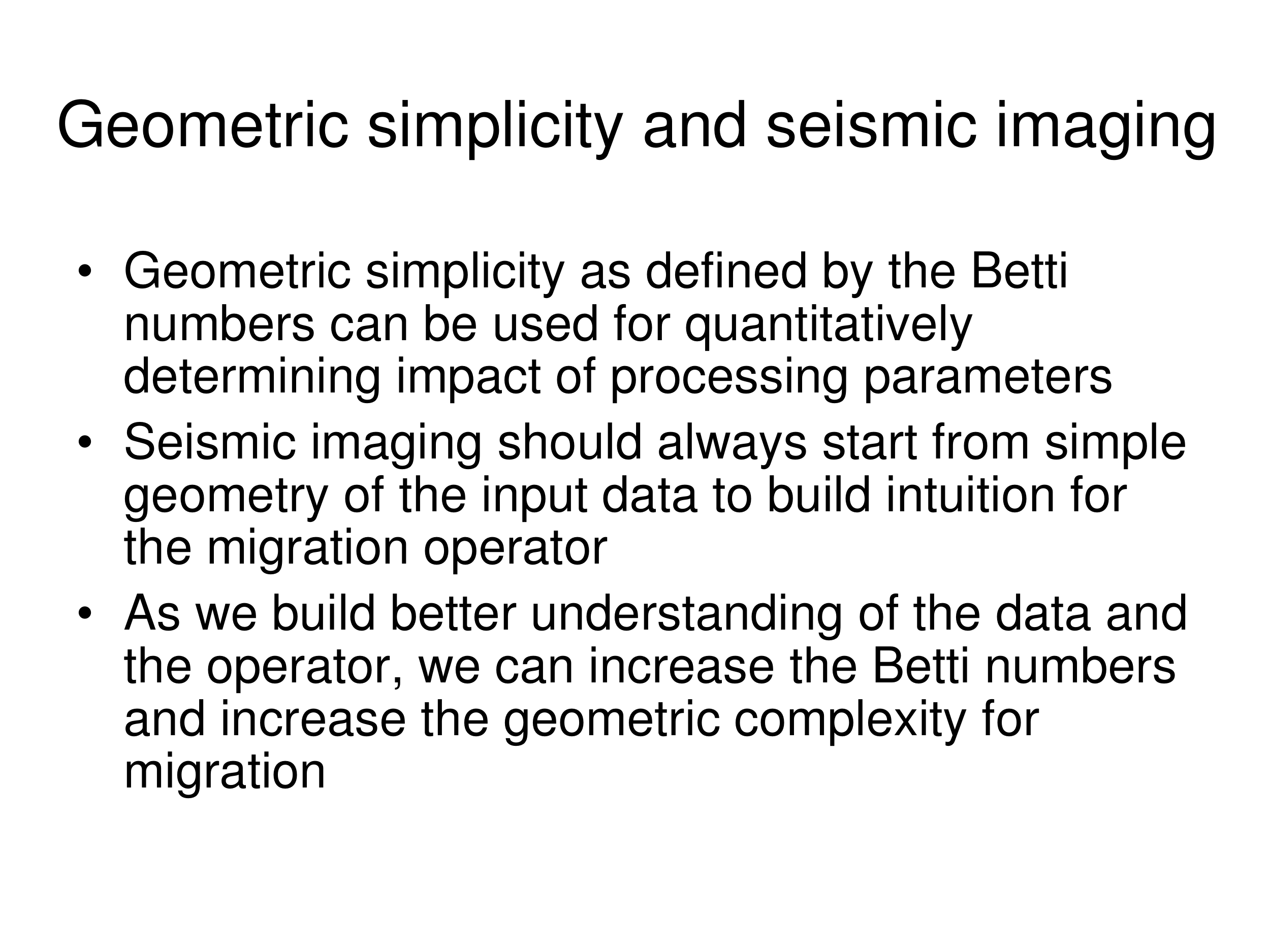}
\caption{Slide 39}
\end{figure}

\begin{figure}
\centering
  \includegraphics[width=5.0in]{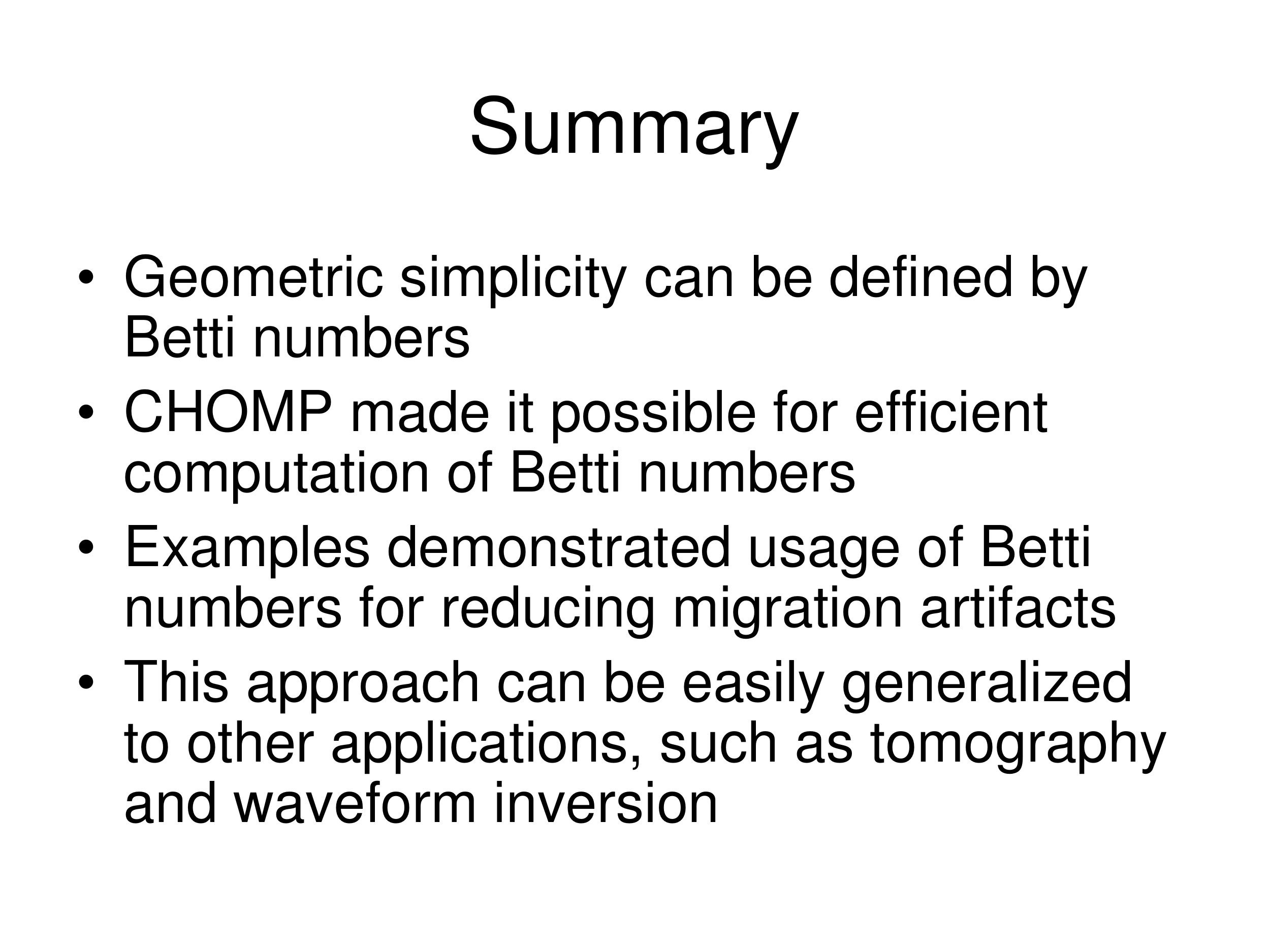}
\caption{Slide 40}
\end{figure}

\clearpage

\begin{figure}
\centering
  \includegraphics[width=5.0in]{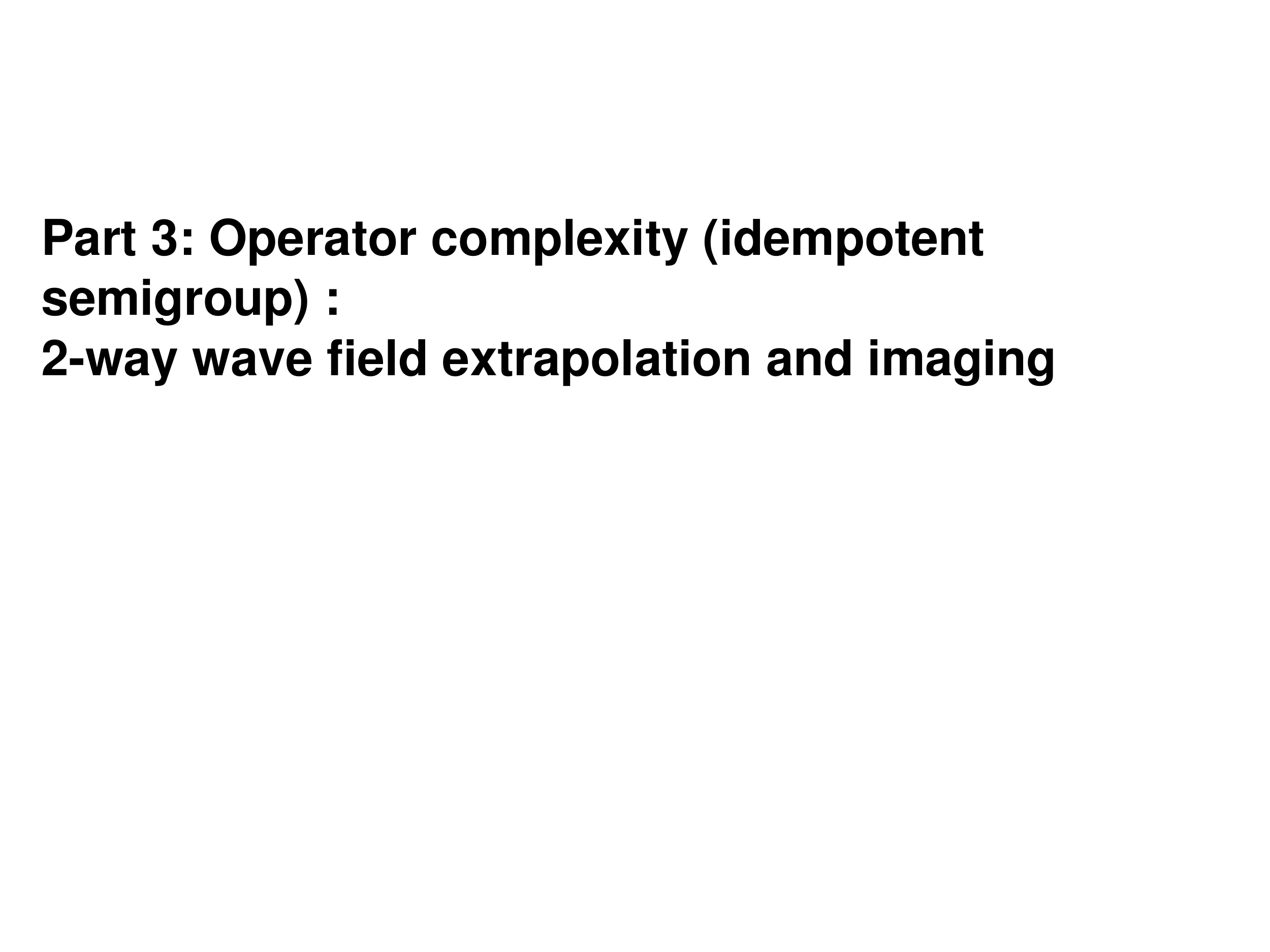}
\caption{Slide 41}
\end{figure}

\begin{figure}
\centering
  \includegraphics[width=5.0in]{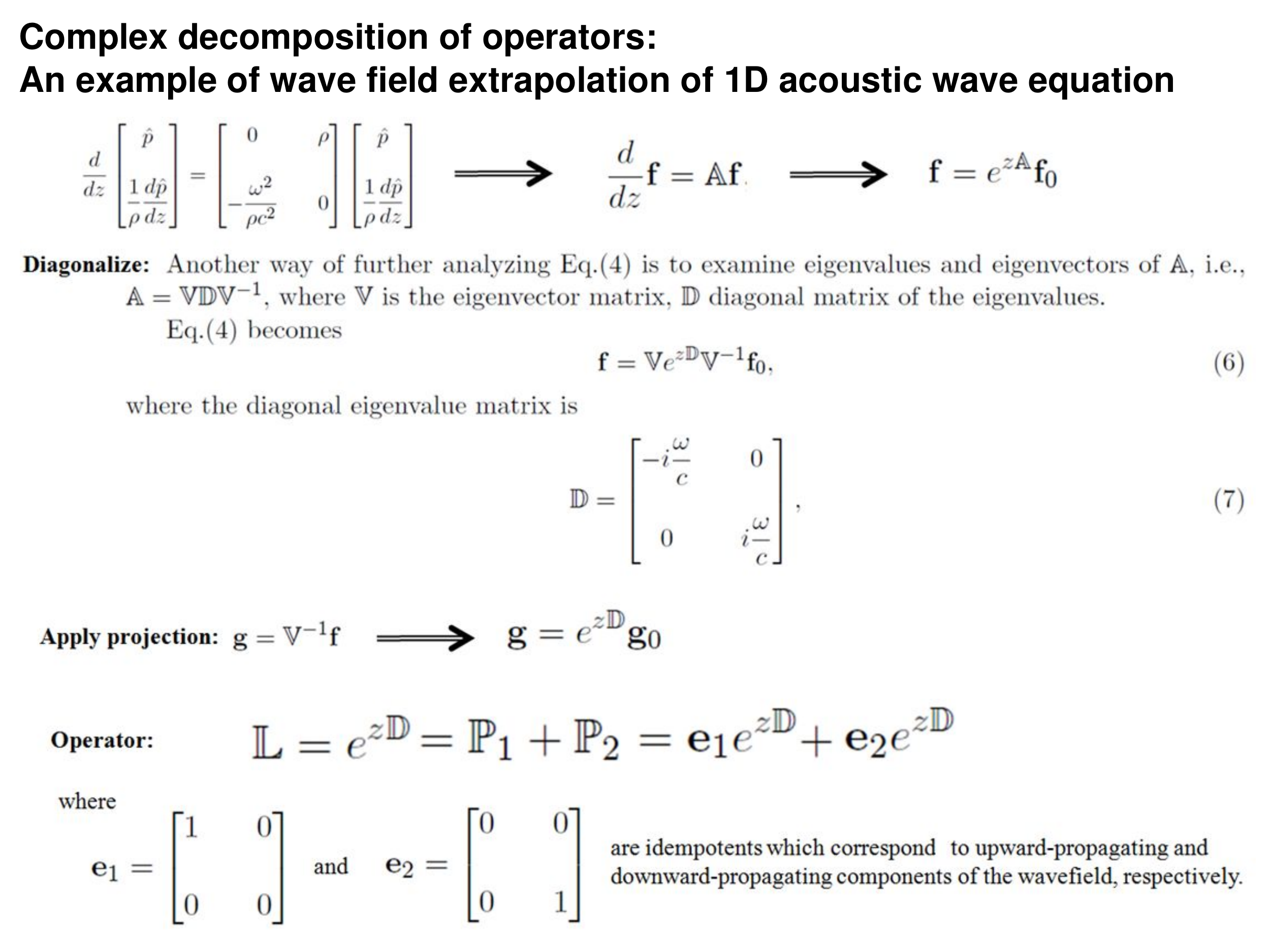}
\caption{Slide 42}
\end{figure}

\begin{figure}
\centering
  \includegraphics[width=5.0in]{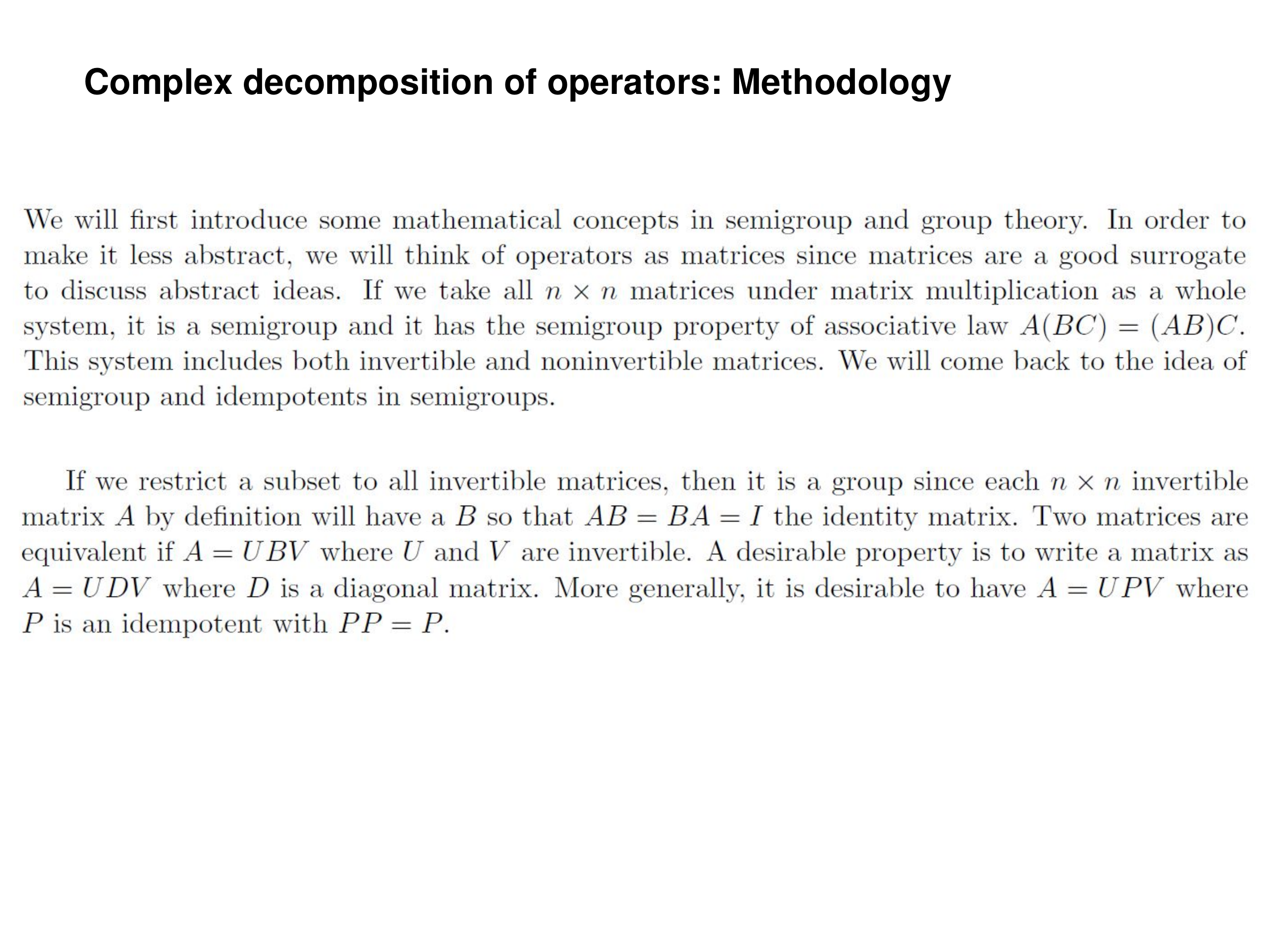}
\caption{Slide 43}
\end{figure}

\begin{figure}
\centering
  \includegraphics[width=5.0in]{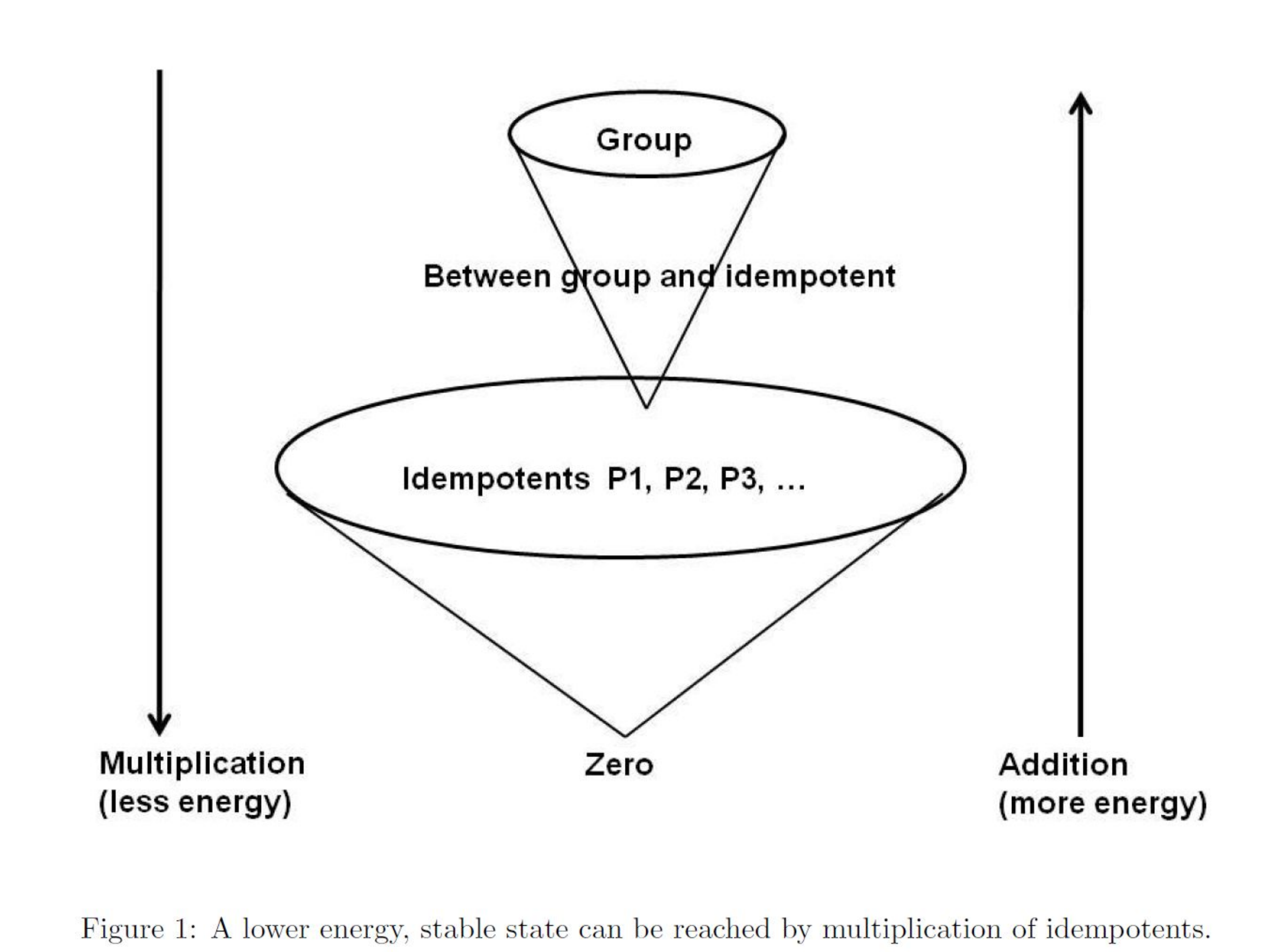}
\caption{Slide 44}
\end{figure}

\begin{figure}
\centering
  \includegraphics[width=5.0in]{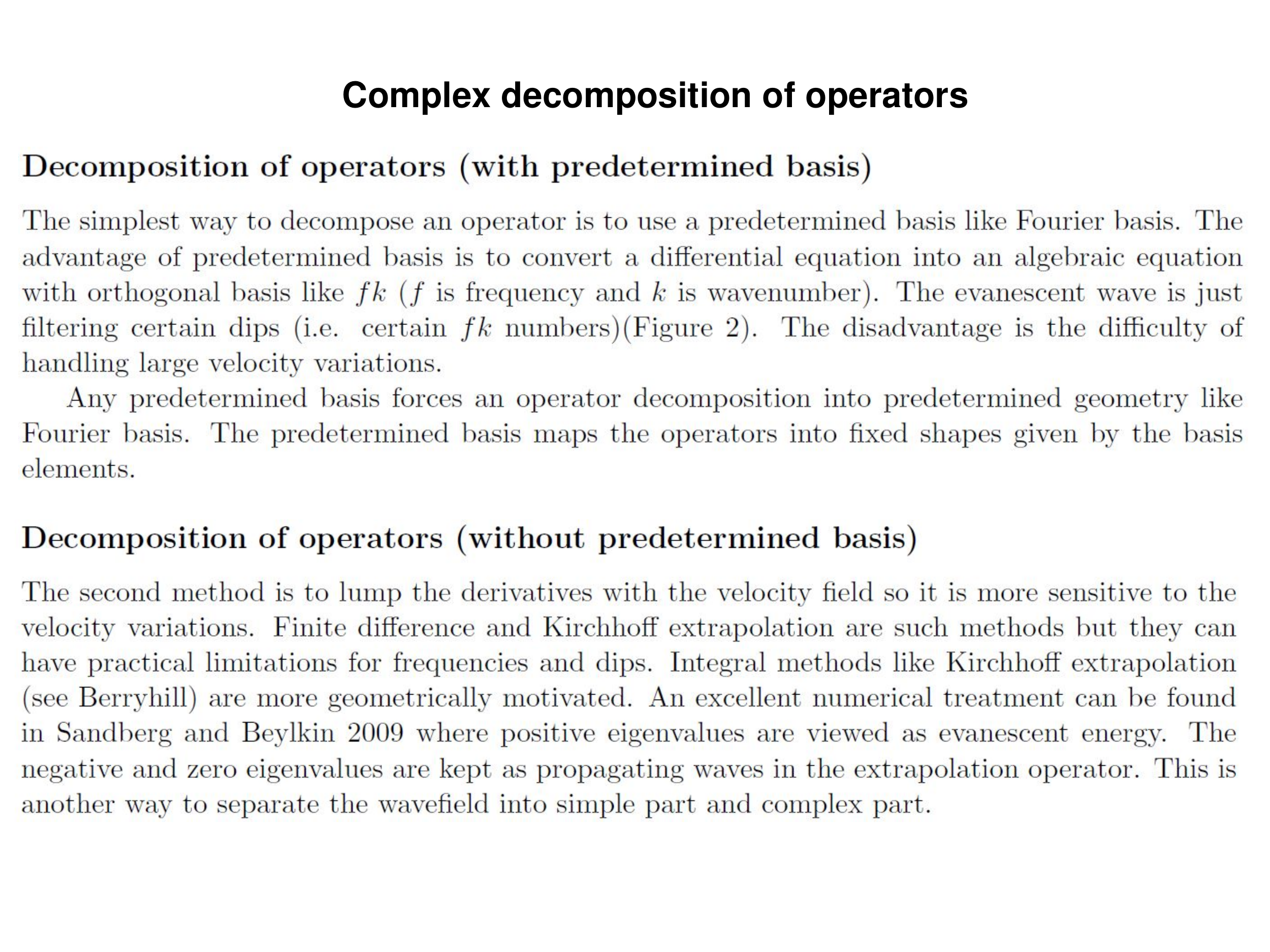}
\caption{Slide 45}
\end{figure}

\begin{figure}
\centering
  \includegraphics[width=5.0in]{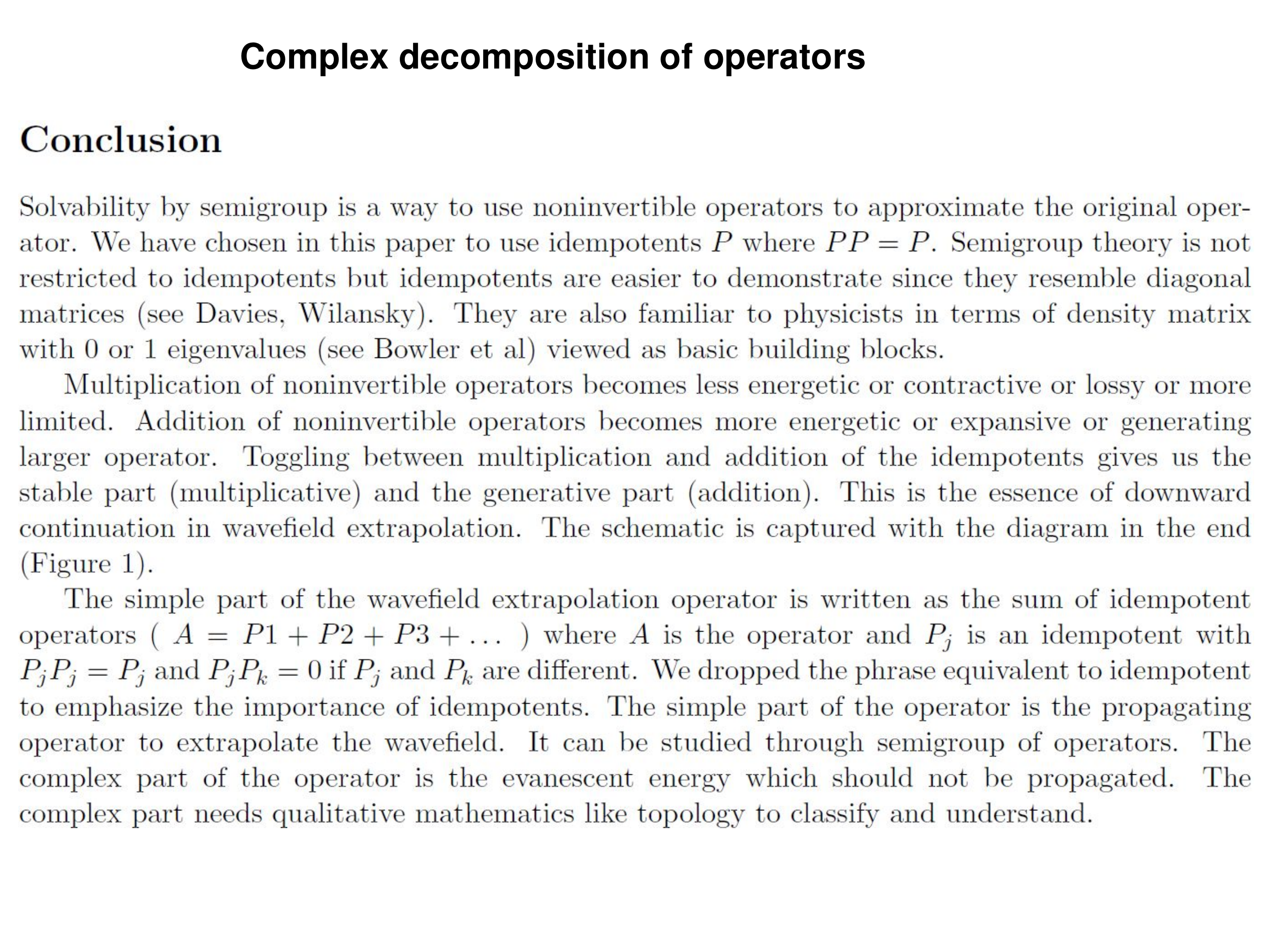}
\caption{Slide 46}
\end{figure}

\begin{figure}
\centering
  \includegraphics[width=5.0in]{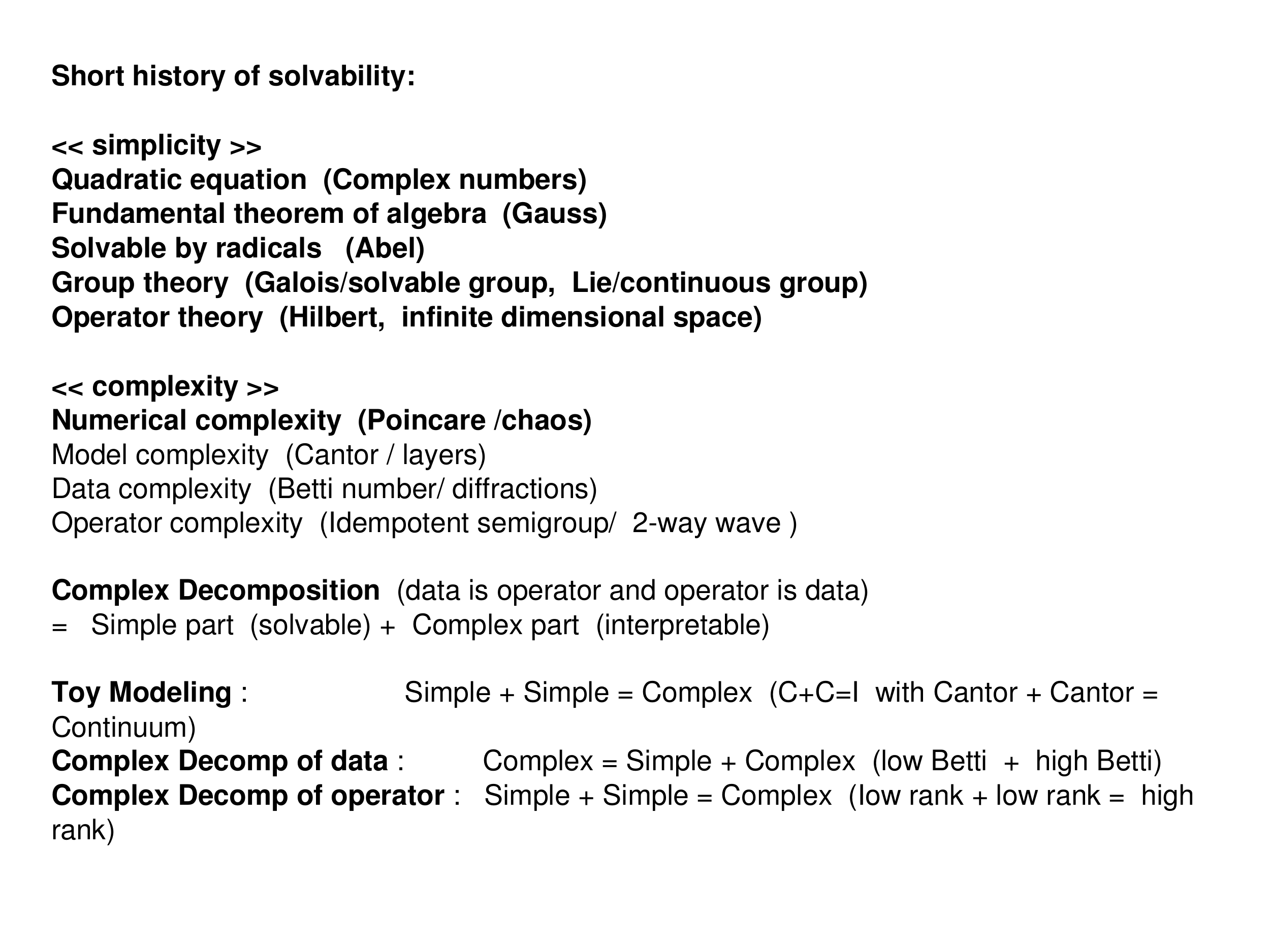}
\caption{Slide 47}
\end{figure}

\end{document}